\documentclass[twocolumn]{aastex63}
\usepackage{natbib}
\usepackage{epsfig}
\usepackage{amsmath}
\usepackage[utf8]{inputenc}
\usepackage{graphicx}  % insert figures
\usepackage{float}  % set the float position of figures
\usepackage{subfigure}  % use subfigures to show the multi figures
\usepackage{booktabs}
\usepackage{xcolor}

\turnoffeditone

\shortauthors{J. Xu et al.}
\shorttitle{CHANG-ES S-band study of NGC~3556}

\begin{document}

\title{CHANG-ES {XXXV}: Cosmic Ray Transport and Magnetic Field Structure of NGC~3556 at 3~GHz}

\author[0009-0006-3887-8988]{Jianghui Xu}
\affiliation{%CAS Key Laboratory for Research in Galaxies and Cosmology, 
Department of Astronomy, University of Science and Technology of China, Hefei, Anhui 230026, People's Republic of China}
\affiliation{School of Astronomy and Space Science, University of Science and Technology of China, Hefei 230026, People's Republic of China}

\author[0000-0001-7254-219X]{Yang Yang}
\affiliation{Purple Mountain Observatory, Chinese Academy of Sciences, 10 Yuanhua Road, Nanjing 210023, China}

\author[0000-0001-6239-3821]{Jiang-Tao Li}
\affiliation{Purple Mountain Observatory, Chinese Academy of Sciences, 10 Yuanhua Road, Nanjing 210023, China}

\author[0000-0003-4286-5187]{Guilin Liu}
\affiliation{%CAS Key Laboratory for Research in Galaxies and Cosmology, 
Department of Astronomy, University of Science and Technology of China, Hefei, Anhui 230026, People's Republic of China}
\affiliation{School of Astronomy and Space Science, University of Science and Technology of China, Hefei 230026, People's Republic of China}

\author[0000-0003-0073-0903]{Judith Irwin}
\affiliation{Dept. of Physics, Engineering Physics \& Astronomy,
Queen's University, Kingston, K7L 3N6, Canada} 

\author[0000-0001-8206-5956]{Ralf-J\"urgen Dettmar}
\affiliation
{Ruhr University Bochum, Faculty of Physics and Astronomy, Astronomical Institute (AIRUB), 44780 Bochum, Germany}

\author[0000-0001-8428-7085]{Michael Stein}
\affiliation
{Ruhr University Bochum, Faculty of Physics and Astronomy, Astronomical Institute (AIRUB), 44780 Bochum, Germany}

\author[0000-0002-3502-4833]{Theresa Wiegert}
\affiliation{Institute of Astrophysics of Andalucía (IAA-CSIC), Glorieta de la Astronomía s/n, 18008 Granada, Spain}

\author[0000-0002-9279-4041]{Q. Daniel Wang}
\affiliation{Department of Astronomy, University of Massachusetts, Amherst, MA 01003, USA}

\author[0000-0001-5310-1022]{Jayanne English}
\affiliation{Department of Physics \& Astronomy, University of Manitoba, Winnipeg, Manitoba, R3T 2N2, Canada}

\correspondingauthor{Jiang-Tao Li}
\email{pandataotao@gmail.com}
\correspondingauthor{Guilin Liu}
\email{glliu@ustc.edu.cn}

\begin{abstract}

Radio halos of edge-on galaxies are crucial for investigating cosmic ray propagation and magnetic field structures in galactic environments. We present VLA C-configuration S-band (\edit1{2--4}~GHz) observations of the spiral galaxy NGC~3556, a target from the Continuum Halos in Nearby Galaxies - an EVLA Survey (CHANG-ES). We estimate the thermal contribution to the radio emission from a combination of the H$\alpha$ and mid-IR data, and employ Rotation Measure Synthesis to reveal the magnetic field structures. In our data, NGC~3556 exhibits a box-like radio halo extending nearly 7~kpc from the galactic plane. The scale height of the total S-band intensity in the halo is $1.68\pm 0.29$~kpc, while that of the non-thermal intensity is $1.93\pm 0.28$~kpc. Fitting the data to a 1-D cosmic-ray transport model, we find advection to describe the cosmic-ray propagation within the halo better than diffusion, with advection speeds of $245 \pm 15$~km~s$^{-1}$ and $205 \pm 25$~km~s$^{-1}$ above and below the disk, respectively. The magnetic field is detected patchily across the galaxy, displaying a toroidal configuration in the rotation measure map. The mean equipartition magnetic field strength is approximately $8.3\ \mu$G in the disk and $4.5\ \mu$G in the halo. In addition, a bubble-like structure extends nearly 3~kpc into the southern halo, aligned with the polarized intensity and H$\alpha$ image, suggestive of superwinds generated by recent star formation feedback in the nuclear region.

\end{abstract}


\section{Introduction} \label{sec:introduction}
The galactic halo is an essential component of a spiral galaxy, consisting of multi-phase gas, dust, cosmic rays (CRs), and magnetic fields (e.g., \citealt{Tumlinson17,Irwin19,Irwin24}). Radio emission from galactic halos is mainly synchrotron radiation of GeV CR electrons in galactic-scale magnetic fields. Recent works find the radio halos of many edge-on spiral galaxies to extend thousands of parsecs above and/or below the disk (e.g., \citealt{Wiegert15,Krause18}). Galactic-scale superwinds, generated by feedback from stellar sources and supermassive black holes, is a major mechanism for producing radio halos (e.g., \citealt{Strickland09,Li16,Li18,Zeng23}). These superwinds, driven or significantly affected by CRs and magnetic fields, play a marked role in regulating CR transport and magnetic field structures (e.g., \citealt{MoraPartiarroyo19a,Schmidt19,Heald22}). Hence, probing radio halos is crucial in understanding the coevolution of CRs and magnetic fields in the circum-galactic medium (CGM) with the galaxy itself.

As significant as the role that CRs and magnetic fields play in galaxy evolution, the knowledge on CR propagation and the global tomography of large-scale magnetic fields remains limited. Non-thermal synchrotron emission in GHz bands is an ideal probe of CRs and magnetic fields, as it dominates thermal emission in this range (e.g. \citealt{Condon92,Irwin12a}). Scrutinizing the spatial variation of non-thermal radio intensity and the non-thermal radio spectral index provides insights into the transportation and radiative loss of CR electrons (e.g. \citealt{Stein19,Heald22}). Moreover, polarization measurements in radio bands, when combined with rotation measure (RM) synthesis, allow for reconstruction of the 3D structure of magnetic fields (e.g., \citealt{MoraPartiarroyo19a,Stein20}).

NGC~3556 is part of the Continuum Halos in Nearby Galaxies - an EVLA Survey (CHANG-ES) sample, a survey of 35 nearby edge-on disk galaxies observed by the Karl G. Jansky Very Large Array (VLA; named the Expanded Very Large Array or EVLA on the approval of the program) during its commissioning phase following the major upgrade in 2012 \citep{Irwin12a,Irwin12b}. NGC~3556 is a normal star-forming galaxy with a star formation rate of $\rm SFR=2.17~M_\odot~yr^{-1}$, and is located at a distance of $d=14.09\rm~Mpc$ \citep{Wiegert15}.
Based on the CHANG-ES C-band (6~GHz) and L-band (1.5~GHz) data, \citet{Miskolczi119} find evidence of a patchy quadrupole magnetic field and CR-driven galactic winds in this galaxy. However, the strong Faraday depolarization in L-band and the low resolution in Faraday space in C-band do not allow for understanding the magnetic field structure. To overcome this limitation, new CHANG-ES observations were taken in S-band with frequency centred at 3~GHz, where the key transition from the Faraday-thin to the Faraday-thick regime takes place. Thus, the combination of our new CHANG-ES data with previous observations (L and C-bands) facilitates unprecedented resolution in Faraday space. 

We present our early results based solely on the S-band data of NGC~3556 in this paper, which is organized as follows: we describe the VLA observations as well as the data reduction procedure in \S\ref{sec:observations}; we present our results in \S\ref{sec:Results}, including polarization and spectral-index measurements. The discussion and conclusion are given in \S\ref{sec:Discussions} and \ref{sec:Summary}, respectively. Measurement uncertainties are reported at the $1\sigma$ confidence levels throughout the paper, unless otherwise noted.

\section{Observations and Data Calibration} \label{sec:observations}

% \begin{table*}
% \begin{center}
% \caption{Information of the \emph{VLA} observations of NGC~3556} 
% \scriptsize
% \begin{tabular*}{0.85\linewidth}{lccccccc}
% \hline\hline
% Date & Integration Time & Configuration & Band & Primary Cal. & Secondary Cal. &Pol Leakage Cal.& SB ID  \\

% \hline
%  13-Jul-2021& 1h 30min&C & S& 3C286& J1035+5628& J1407+2827& 39738602\\
%  23-Jul-2021& 1h 30min&C & S& 3C286& J1035+5628& J1407+2827& 39738321\\ 
% \hline
% \end{tabular*}
% \label{table:VLAObs}
% \end{center}
% \end{table*}

\subsection{CHANG-ES VLA S-band data calibration and analysis} \label{subsec:VLAObservations}

\subsubsection{VLA S-band observations} \label{subsubsec:VLAObservations}

\edit1{The original CHANG-ES survey focused on observations in the C-band and L-band \citep{Irwin12a}. The S-band (2--4 GHz; central frequency 3~GHz) survey is a follow-up program approved for the 2020A, 2021A, and 2022B semesters in C-configuration, with a total exposure time of about 118~hours for 34 of the 35 CHANG-ES galaxies (PI: Y. Stein).} 
NGC~3556 was observed on July 13 and 23, 2021, with 27 antennas in C-configuration. The on-source time per day is about 1.5~hours, \edit1{for} 3 hours in total. The S-band with a bandwidth of 2~GHz is divided into 16 spectral windows, of which two are initially flagged due to the existence of significant radio frequency interference (RFI). In addition, we remove other unusable low-quality data, resulting in a final effective bandwidth of about 1.2~GHz.

We used version 6.2.1 of the Common Astronomy Software Application package \edit1{\citep[CASA,][]{CASA22}} to calibrate and visualize the data. For both data sets, 3C 286 acts as the primary calibrator that serves the purpose of determining the bandpass and the absolute polarization angle. The secondary calibrator is J1035+5628, and the zero-polarization calibrator is J1407+2827. \edit1{Before the final imaging process}, self-calibration is performed only once in the phase. Further attempts of self-calibration yield minimal improvement in the quality of the final datasets. \edit1{Finally, Stokes $I$ map is imaged using a robust parameter of zero, with deconvolution performed using the Multi-Scale Multi-Frequency Synthesis algorithm \citep{mtmfs11}.} \edit1{The output images are presented at a reference frequency of 3~GHz.}

% We use the Stokes $Q$ and $U$ maps to extract information of polarization and magnetic fields. The linearly polarized intensity image is obtained by using the relationship $P=\sqrt{Q^{2}+U^{2}-\sigma_{Q,U}^{2}}$, where the $\sigma_{Q,U}$ is the r.m.s. noise of the Stokes $Q$ and $U$ maps. The polarization angle of the observed electric field vector, denoted as $\chi$, is calculated as $\chi=1/2 \arctan (U/Q)$. 
% The quantity $\chi+90^\circ$ represents the apparent orientation of the magnetic field in the sky plane.
% In addition, we perform RM synthesis to estimate the strength of the magnetic field perpendicular to the plane of the sky, of which the details are deferred to the following section.

The r.m.s. noise of the resultant Stokes $I$ map reaches $4.8 \mathrm{\ \mu Jy\ beam}^{-1}$, with a beam size of $5.68''\times 4.50''$ (Fig.~\ref{fig:total_intensity}a). To highlight the structures on larger scales revealed at a lower resolution, we also generate a uv-tapered version amidst the cleaning process using a setting parameter of \texttt{'10 arcsec'}, which leads to an r.m.s. noise of $5.5 \mathrm{\ \mu Jy\ beam}^{-1}$ and a beam size of $10.47''\times 10.15''$ (Fig.~\ref{fig:total_intensity}b).

\subsubsection{Rotation Measure synthesis} \label{subsubsec:RMsynthesis}

When linearly polarized radiation from a source propagates through a magnetized plasma, such as the interstellar medium (ISM) of galaxies, its polarization angle undergoes wavelength-dependent rotation due to the birefringence of the magneto-ionic medium \edit1{(the rotation angle $\beta={\rm RM}~\lambda^2$, where $\lambda$ is the wavelength)}. This phenomenon is known as the Faraday rotation. In this work, we employ the RM synthesis technique \citep{Burn96,RMsyn} to determine the intrinsic polarization angle and analyze the line-of-sight component of the magnetic field. This technique establishes a relationship between the complex polarized surface brightness in Faraday depth space and $\lambda^{2}$-space through a Fourier transform.

To perform RM synthesis in the S-band, we initially divided each of the 14 available spectral windows into four segments, each consisting of 13 channels due to the common flagging of the edge channels 0--5 and 58--63. As a result, a frequency bandwidth of 26 MHz is achieved for each individual segment, corresponding to $1.73\mathrm{\ cm}^{2}$ in $\lambda^{2}$-space. The parameter setting for the RM synthesis is detailed in Table \ref{tab:RMsyn}. 

\begin{table}
    \begin{center}
        \caption{Rotation Measure synthesis parameters}
        \begin{tabular*}{0.8\columnwidth}{p{4cm}c}
            \hline\hline
            & S-band \\
            \hline
            $\nu_{\rm min}$ (GHz)& 2\\
            $\nu_{\rm max}$ (GHz)& 4\\
            $\Delta \lambda^2$ (m$^2$)& $1.7\times10^{-2}$\\
            $\delta \phi\ (\mathrm{rad\ m}^{-2})$ & $2.0\times10^{2}$\\
            \hline
            $\lambda_{\rm min}$ (m) & $7.5\times10^{-2}$\\
            $\mathrm{max}_{\mathrm{scale}}\ (\mathrm{rad\ m}^{-2})$ & $5.6\times10^2$\\
            \hline
            $\delta \nu$ (MHz) & 26\\
            $\delta \lambda^2$ (m$^2$) & $1.7\times10^{-4}$\\
            $|| \phi_{\mathrm{max}}||\ (\mathrm{rad\ m}^{-2})$ & $1.0\times10^4$\\
            \hline
        \end{tabular*}
        \label{tab:RMsyn}
    \end{center}
\footnotesize \textbf{Notes:} $\nu_{\rm min}$ and $\nu_{\rm max}$, minimum and maximum frequencies of S-band; $\Delta \lambda^2$, the width of the $\lambda^2$ distribution; $\delta \phi\approx2\sqrt{3}/\Delta \lambda^2$, FWHM resolution in Faraday depth space; $\mathrm{max}_{\mathrm{scale}} \approx \pi / \lambda^2_{\rm min}$, the largest continuous Faraday depth range detectable in a given observation; $\delta \nu$, the channel width; $\delta \lambda^2$, the channel width in $\lambda^2$-space\edit1{, calculated at a frequency of 3~GHz}; $|| \phi_{\mathrm{max}}|| \approx \sqrt{3}/\delta \lambda^2$, maximum observable RM.
\end{table}

For each of these segments, the Stokes $Q$ and $U$ images were produced separately using \edit1{Briggs} weighting with a robust parameter of 2. This weighting scheme optimizes the sensitivity to recover extended faint emission without uv-tapering. All resulting maps are convolved to a uniform beam size of $15'' \times 15''$ \edit1{before applying} the primary-beam correction, and are then consolidated into two individual datacubes for Stokes $Q$ and $U$. RM synthesis is conducted using the \texttt{RM-Tools} code \citep{RMtools}, employing a Faraday depth ($\phi$) ranging from $-2048 \mathrm{\ rad\ m}^{-2}$ to $+2048 \mathrm{\ rad\ m}^{-2}$ and a sample spacing of $\delta \phi = 4 \mathrm{\ rad\ m}^{-2}$. This procedure results in $Q$, $U$, and PI (polarized intensity) Faraday spectra, where two image axes and one Faraday depth axis are covered. These Faraday spectra were subsequently cleaned down to a 3$\sigma$ noise level of the uncleaned $Q$ and $U$ spectra.

Finally, the PI value is determined by fitting the peak value along the Faraday depth axis. \edit1{The peak is identified by fitting a parabola to the highest PI value and the two adjacent points in the Faraday spectrum.} \edit1{A} Ricean bias correction is \edit1{then} applied \edit1{ using the formula $P_{\rm eff}=\sqrt{P^{2}-2.3\sigma_{\rm FDF}^2}$, where $\sigma_{\rm FDF}$ represents the estimated noise in the PI Faraday spectrum}. The corresponding RM value is the best-fit $\phi$ of the peak. The Stokes values $Q$ and $U$ for the peak are derived by interpolating between the two samples closest to the fitted peak. The intrinsic polarization angle (PA) is calculated based on the aforementioned $Q$ and $U$ values and then corrected using the RM value and the central wavelength \edit1{($\lambda^{2}=1.03\times10^{-2}{\rm\ m}^2$, correspond to a reference frequency of 2.95~GHz)}. We apply this strategy to each pixel of the image axes to create maps for PI, PA, and RM, along with their uncertainty maps.

We note that the RM map presented here demonstrates the most prominent peak position of each pixel only, without characterizing other line-of-sight components. In our final RM map, the Galactic foreground RM in the direction of NGC~3556 ($\mathrm{RM}_{\mathrm{fg}}=11.0\pm 2.2 \mathrm{\ rad\ m}^{-2}$) has been subtracted, using the data from \citet{Hutschenreuter22}.

\subsection{Other multi-wavelength data} \label{subsec:multiwavelengthdata}

The comparison of our S-band data to those obtained at other wavelengths is informative. In particular, the H$\alpha$ and mid-IR data allow for the estimation of the thermal fraction (originating from star formation) in galactic radio emission (\S\ref{subsec:ThermalNonthermalMap}). We also use the low-frequency radio data to better constrain the non-thermal radio emission (\S\ref{subsec:ThermalNonthermalMap}). 

We use the H$\alpha$ image from the Apache Point Observatory (APO) 3.5m telescope equipped with the Astrophysical Research Consortium Telescope Imaging Camera (ARCTIC). The size of the pixels is $0.228''$ and the [\ion{N}{2}] line contamination is 0.20. Data reduction is described in \citet{Vargas19}.

The archival \edit1{\textit{WISE}} 22$\mu$m data from \citet{Wright10} are also used. The 22$\mu$m fluxes are multiplied by a factor of 1.03 to convert to 24$\mu$m fluxes, based on the tight linear relationship between them, as reported by \citet{Wiegert15}. To complete the thermal/nonthermal decomposition, we downgrade the H$\alpha$ map to an angular resolution of $15''$(FWHM) by convolving it with a Gaussian kernel. The \edit1{\textit{WISE}} 22$\mu$m map is smoothed to the same resolution using the kernel provided by \citet{Aniano11}. These two maps are then re-gridded to match our S-band radio maps.

Low-frequency radio data are critical to constrain the slope and curvature of synchrotron emission, and the LOFAR 144~MHz image from the data release 2 \citep[DR2,][]{Heesen22} of the LOFAR Two-meter Sky Survey \citep[LoTSS,][]{Shimwell17} is used for this purpose. The DR2 images are available at resolutions of $6''$ and $20''$. We choose the $6''$ image for NGC~3556 and downgrade it to $15''$ to facilitate comparison with the S-band data. 

\section{Results} \label{sec:Results}

\subsection{Total intensity of radio continuum} \label{subsec:Intensity}

\begin{figure*}[th]
    \begin{center}
        \includegraphics[width=0.95\linewidth]{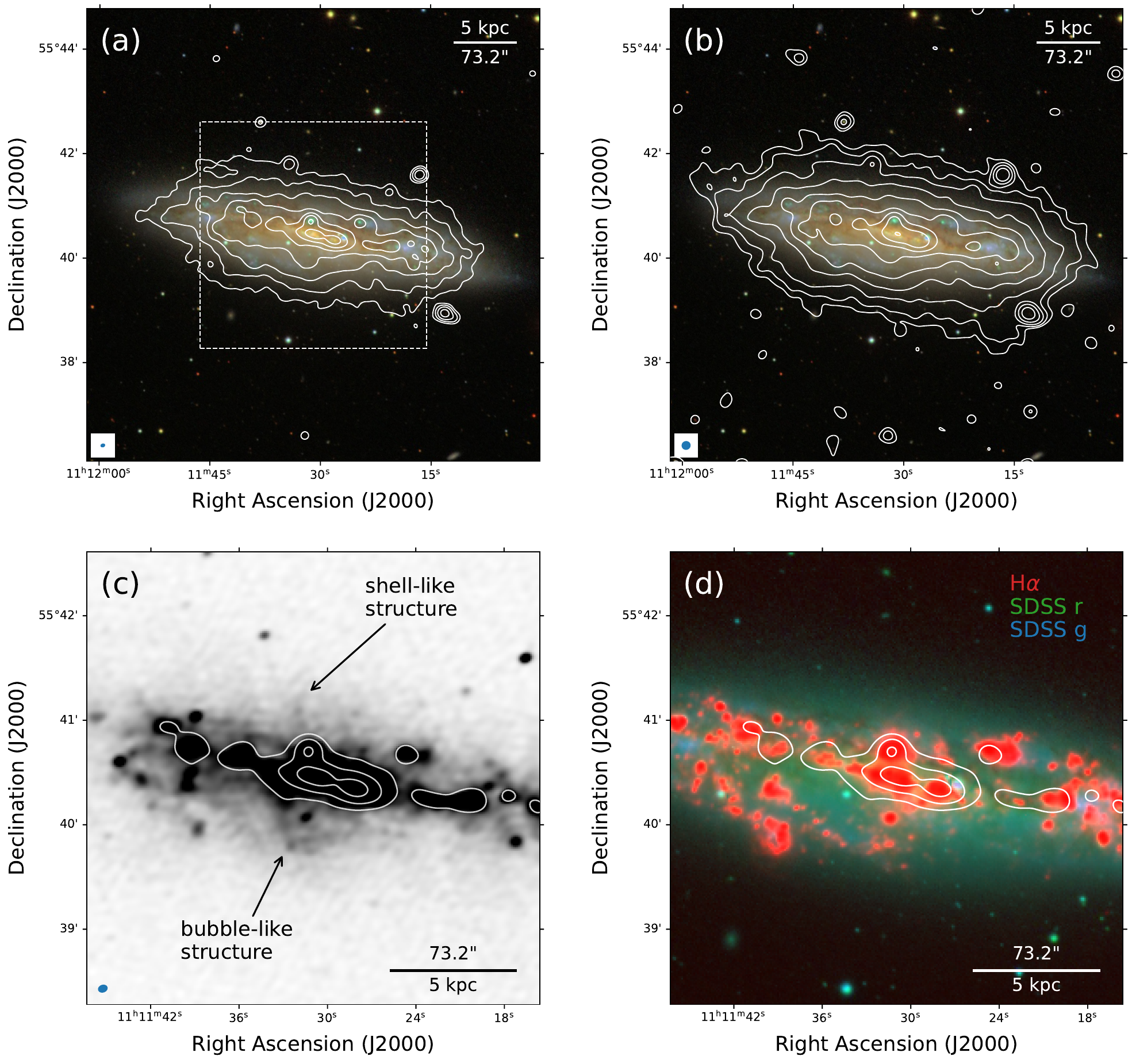}
        \caption{S-band total intensity images of NGC~3556 compared to multi-wavelength images. (a) SDSS optical image made from the \textit{i, r, g} filters. The white contours represent the image with robust zero weighting, ranging from $3 \sigma = 14.3\ \mu\mathrm{Jy\ beam}^{-1}$ and increasing by a factor of two. The beam size for this panel is $5.68''\times 4.50''$. The white box delineates the area depicted in panels (c) and (d). (b) $10''$ uv-tapered image superimposed on the SDSS optical image. Contours start at $3 \sigma = 16.5\ \mu\mathrm{Jy\ beam}^{-1}$ and are in increments of 2. The beam size for this panel is $10.47''\times 10.15''$. (c) Robust zero weighting image displayed in grayscale. The white contours at $48, 96, 192 \sigma$ levels represent the inner region. Noteworthy features include a bubble-like structure in the southern halo and a shell-like structure in the northern halo, as indicated by arrows. (d) Tri-color image made from the H$\alpha$ and SDSS \textit{r, g} filters overlaid with the same contours as in panel (c). The beam is displayed as a blue ellipse in the bottom-left corner of each panel.}
        \label{fig:total_intensity}
    \end{center}
\end{figure*}

The S-band total intensity maps of NGC~3556 are presented in Fig.~\ref{fig:total_intensity}. When robust zero weighting is used (Fig.~\ref{fig:total_intensity}a), the morphology of the radio halo is a well-defined box-like structure, extending up to $5.1\mathrm{\ kpc}$ above and below the galactic plane, as depicted by the $3\sigma$ contour. The total flux density measured within this contour is $150\pm 5 \mathrm{\ mJy}$. In the uv-tapered total intensity map (Fig.~\ref{fig:total_intensity}b), the radio halo retains its box-like shape, although it is slightly more extended along the disk on the eastern side. The vertical extent of the halo increases to $\sim 6.7 \mathrm{\ kpc}$, and the total flux density enclosed by the $3\sigma$ contour is $155 \pm 5 \mathrm{\ mJy}$.

We zoom in on the S-band total intensity map and compare the small structures therein with optical images in Fig.~\ref{fig:total_intensity}c,d. A general correlation exists between the bright structures in the S-band map and those in the H$\alpha$ image. Moreover, above the central disk, several prominent extended structures can be identified (marked by arrows). One notable feature is a bubble-like structure on the southern side, extending approximately 3~kpc from the galactic plane to the halo. This feature originates from the central region of the galaxy that is bright in H$\alpha$, and spatially coincides with a chain of H$\alpha$ knots off the galactic midplane. On the opposite side, we detect a shell-like structure in the northern halo without an evident counterpart in the H$\alpha$ image. Furthermore, in \S\ref{subsec:Polarization}, we demonstrate that the polarization intensity and magnetic field orientations are well aligned with the bubble-like structure, suggesting that the magnetic fields, along with the CRs, are largely related to a galactic outflow.

\subsection{Thermal/non-thermal separation and spectral index} \label{subsec:ThermalNonthermalMap}

The radio continuum emission originating from star-forming galaxies comprises both thermal free-free emission and nonthermal synchrotron emission \citep{Condon92}. To delineate the magnetic fields and cosmic rays (CRs) associated with the non-thermal emission, it is imperative to effectively decompose the thermal / non-thermal contribution to the total intensity of the radio continuum emission.

In this work, we employ the mixture method developed by \cite{Vargas18} to estimate the thermal contribution, in which a combination of H$\alpha$ and 24 $\mu$m emission is used to create a star formation rate map:

\begin{multline}
    \mathrm{SFR}_{\mathrm{mix}}[\mathrm{M}_{\odot}\  \mathrm{ yr}^{-1}]=5.37\times10^{-42}\cdot (
        L_{\mathrm{H}\alpha\_ \mathrm{obs}}\\
        +0.042\cdot \nu L_{24\mu \mathrm{m}})[\mathrm{erg\ s}^{-1}] \label{equ:SFR}.
\end{multline}

Here, we adopt a new coefficient of 0.042 derived by \cite{Vargas18}, where a correction factor of 1.36 has been applied to the coefficient given by \cite{Calzetti07} to account for edge-on galaxies. Hence, the thermal emission is calculated directly from the SFR \citep{Murphy11}:

\begin{multline}
    L_{\nu}^T[\mathrm{erg\ s}^{-1}\ \mathrm{Hz}^{-1}] = 2.2\times 10^{27}\left(\frac{T_{\mathrm{e}}}{10^4\mathrm{\ K}}\right)^{0.45} \\
    \times\left(\frac{\nu}{\mathrm{GHz}}\right)^{-0.1}\left(\frac{\mathrm{SFR}_{\mathrm{mix}}}{\mathrm{M}_{\odot}\  \mathrm{ yr}^{-1}}\right) \label{equ:L_thermal}.
\end{multline}

We adopt a typical electron temperature value of $T_{\mathrm{e}}=10^4 \mathrm{\ K}$. Using Equations (\ref{equ:SFR}) and (\ref{equ:L_thermal}), we create thermal emission maps at frequencies of 144 MHz and 3 GHz, which are subsequently subtracted from the corresponding radio maps, so that the non-thermal emission maps are obtained. During this process, we smooth all the relevant maps to a common beam size of $15'' \times 15''$ and regrid them to the same pixel size. Further details of this analysis are given in \cite{Vargas18}.

\begin{figure}[th]
    \begin{center}
        \includegraphics[width=8.5cm]{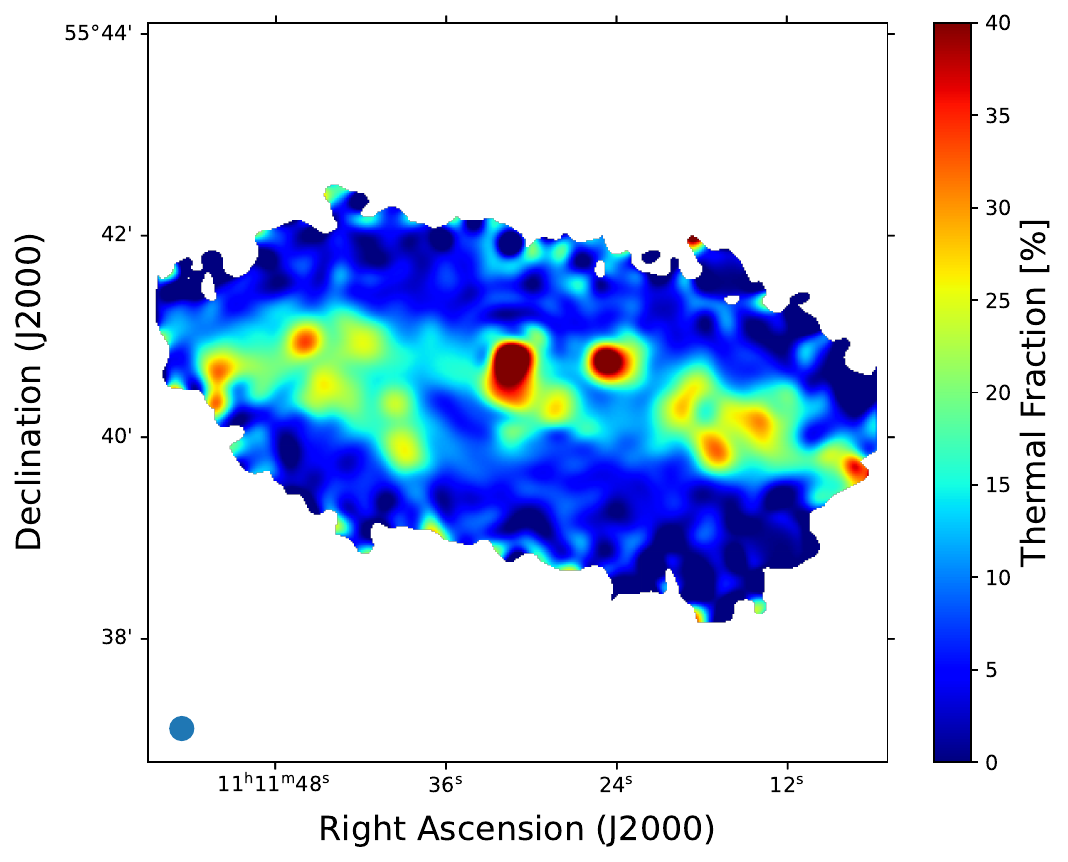}
        \caption{NGC~3556 thermal fraction map at S-band. Only pixels where the total intensity was larger than $3 \sigma$ were taken into account. The beam of $15'' \times 15''$ is displayed as a blue circle in the bottom-left corner.}
        \label{fig:thermal_fraction}
    \end{center}
\end{figure}

The S-band thermal fraction map is shown in Fig. \ref{fig:thermal_fraction}, in which a rapid decline from the galactic disk to the halo region is evident. The median thermal fraction is 19.6\% on the disk, while it exceeds 50\% at the galactic center. In contrast, the thermal fraction is significantly lower in the northern and southern halo, measuring 4.9\% and 4.7\%, respectively. The distribution of thermal emission appears relatively symmetric. At LOFAR 144MHz, the thermal fraction is considerably lower, with a median of 4.2\% and 0.6\% in the disk and halo, respectively. Even in the galactic center, it reaches only 25\% or lower. These findings are consistent with observations of other edge-on galaxies \citep[e.g., ][]{Vargas18, Stein19, MoraPartiarroyo19b}.

\begin{figure}[th]
    \begin{center}
        \includegraphics[width=8.5cm]{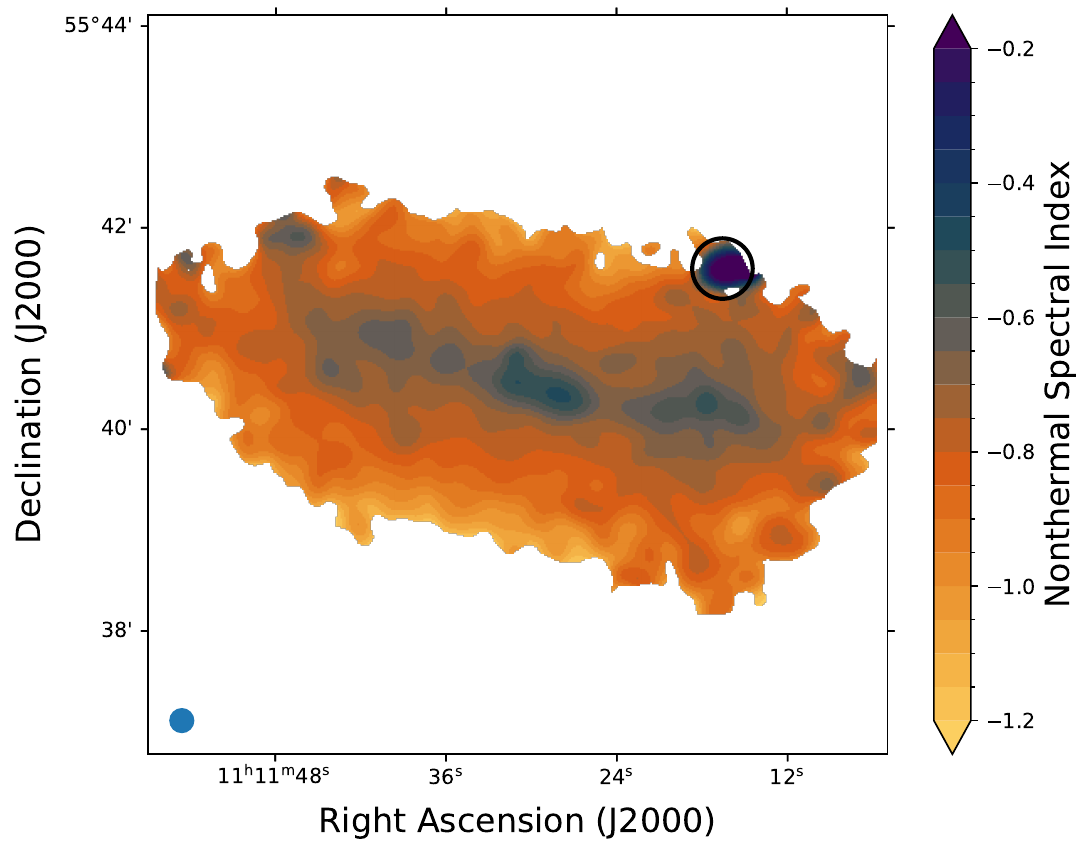}
        \caption{Distribution of the nonthermal spectral index calculated between S-band and LOFAR, visualized using the colormap from \citet{colormap}. The point source with a flat spectral index at $-0.2$ is marked with a black circle. The beam of $15'' \times 15''$ is displayed as a blue circle in the bottom-left corner.}
        \label{fig:spectral_index}
    \end{center}
\end{figure}

Using the nonthermal maps obtained in the S-band and at LOFAR 144 MHz, we create the nonthermal spectral index map by performing the following calculation:

\begin{equation}
    \alpha_{\mathrm{nth}}=\frac{\log [I_{\mathrm{nth}}(\nu_{1})/I_{\mathrm{nth}}(\nu_{2})]}{\log (\nu_{1}/\nu_{2})}.
\end{equation}
Here, $\nu_{1}$ and $\nu_{2}$ represent the frequencies of 3 GHz and 144 MHz, respectively. The resulting spatial distribution of the nonthermal spectral index is shown in Figure \ref{fig:spectral_index}.

In the galactic disk, we find a nonthermal spectral index of $-0.64\pm0.02$; while in the central region, where H$\alpha$ emission is prominent, this index reaches as low as $-0.50\pm0.04$. These values are consistent with the expected injection spectral index of cosmic ray energy spectra originating from supernova remnants, where the cosmic rays are generated \citep{Bell78,Biermann93}. 
Away from the disk, the spectrum steepens, with the mean spectral index in the halo being around $-0.85\pm0.03$, and even lower in the outermost region. This steepening is consistent with the expected aging of cosmic-ray electrons.

Furthermore, in the northwest region, approximately 6 kpc away from the disk, an anomalous point source with a remarkably flat spectral index of $-0.2\pm0.2$ is observed (Source \#7 in \S\ref{subsec:PointSources}). It is intriguing to find that the thermal fraction of this point source is rather low, being less than $1\%$ in the S-band.

\subsection{Vertical scale heights} \label{subsec:ScaleHeight}

To quantify the vertical extent of the radio emission inside the galactic halo, we adopt an exponential profile-fitting approach. Following the method described in \cite{Dumke95}, we assume an intrinsic exponential profile characterized by a peak intensity $w_{0}$ and a scale height $z_{0}$:

\begin{equation}
    w(z) = w_{0}\exp (-z/z_{0}).
\end{equation}

To account for the contribution of the telescope beam and the projected emission of the thin disk, we convolve this profile with the effective telescope beam $\mathrm{HPBW}_{\mathrm{eff}} = 2\sqrt{2\ln 2} \cdot \sigma_{\mathrm{eff}}$, which can be described as a Gaussian function:

\begin{equation} \label{equ:gaussian_beam}
    g(z) = \frac{1}{\sqrt{2\pi \sigma^{2}_{\mathrm{eff}}}}\exp (-z^{2}/2\sigma^{2}_{\mathrm{eff}}).
\end{equation}

The resulting convolved emission profile then takes this form:

\begin{equation} \label{equ:profile_exp}
\begin{aligned}
    W_{i}(z) &= \frac{w_{i}}{2}\exp (-z^{2}/2\sigma^{2}_{\mathrm{eff}})\\
    &\times\left[
        \exp \left( \frac{\sigma^{2}_{\mathrm{eff}} - zz_{i}}{\sqrt{2}\sigma_{\mathrm{eff}}z_{i}} \right)^{2} \mathrm{erfc} \left( \frac{\sigma^{2}_{\mathrm{eff}} - zz_{i}}{\sqrt{2}\sigma_{\mathrm{eff}}z_{i}} \right) \right. \\
    & + \left. \exp\left( \frac{\sigma^{2}_{\mathrm{eff}} + zz_{i}}{\sqrt{2}\sigma_{\mathrm{eff}}z_{i}} \right)^{2} \mathrm{erfc} \left( \frac{\sigma^{2}_{\mathrm{eff}} + zz_{i}}{\sqrt{2}\sigma_{\mathrm{eff}}z_{i}} \right)\right],
\end{aligned}
\end{equation}

where $\mathrm{erfc}$ is the complementary error function, and $i=0$ or $1$ corresponds to the galactic disk and halo components, respectively. 
Applying Eq. \ref{equ:profile_exp} twice allows for decomposing the radio emission into the contribution of the galactic disk and the halo. These components are characterized by their respective amplitudes ($w_{0}$ and $w_{1}$) and scale heights ($z_{0}$ and $z_{1}$). However, due to the complexity of the two-component model, data fits may suffer from large uncertainties in the model parameters or even be infeasible at all. When fits are unsatisfactory (e.g. \edit1{the relative uncertainty in the parameter values exceeds 200\%}), we simply employ a one-component exponential profile. This strategy proves effective in constraining the fitting parameters. 

The vertical scale heights of NGC~3556 are determined using the \texttt{BoxModels} task in the \texttt{NOD3} package \citep{Muller17}, in which multiple strips, each consisting of several rectangular boxes, are selected and juxtaposed perpendicular to the major axis of the galaxy. We average the radio intensity in every box, and the uncertainties are calculated as the standard deviation of the mean. The averaged intensity distribution along each strip is used for fitting to determine the vertical scale heights. 

\begin{figure}[th]
    \begin{center}
        \includegraphics[width=7.5cm]{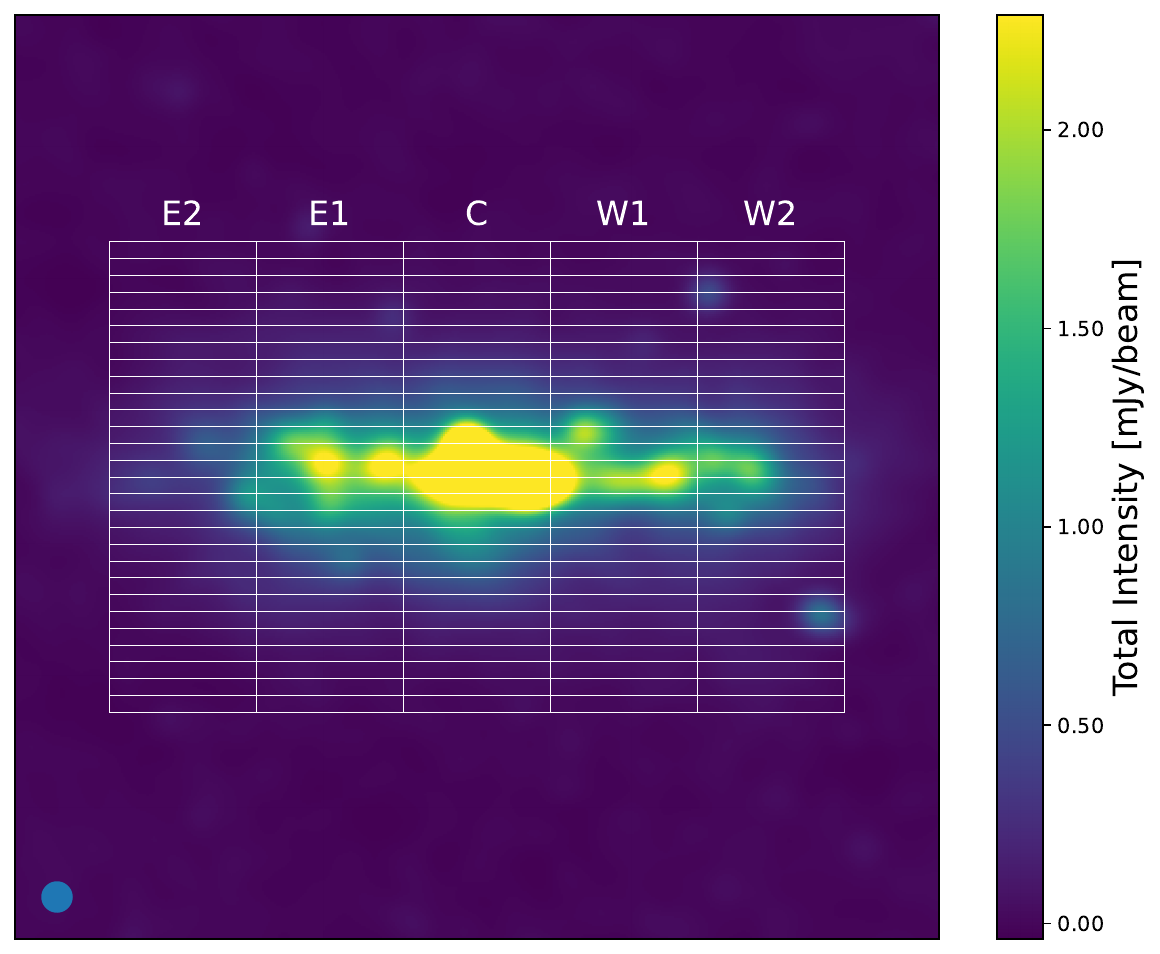}
        \caption{Strip setup on the total intensity map of NGC~3556. The galaxy has been split up into five strips: Eastern 2 (E2), Eastern 1 (E1), Central (C), Western 1 (W1), Western 2 (W2). Individual boxes have a height of $8''$ and a width of $70''$. Upper boxes have a positive offset, and lower boxes have a negative one. The beam size of $15''$ is given in the bottom left corner.}
        \label{fig:strip_setup}
    \end{center}
\end{figure}

We apply the strategy to the uv-tapered map of total intensity and the nonthermal synchrotron intensity map that are generated in \S\ref{subsec:ThermalNonthermalMap}. To facilitate the comparison of them, we smooth the total intensity map to match the $15'' \times 15''$ beam size of the synchrotron map. We divide the galaxy into five strips, of which each has a width of $70''$, to assess the difference between the central and outer regions of the galaxy. We set the height of a box to be $8''$, or about half of the beam width. To guarantee that those regions down to $2 \times \mathrm{rms}$ level are also sampled, we place 28 boxes in each strip. The strip setup is illustrated in Figure \ref{fig:strip_setup}.

In Eq. \ref{equ:gaussian_beam}, the projection correction requires the inclination angle of NGC~3556, $81^\circ$ as given by \cite{Irwin12a}.
However, when fitting the vertical intensity distribution, we find that reasonable results can only be obtained when an inclination angle of $\geq 88^\circ$ is adopted. A similar situation occurred with NGC~4631, as presented in \cite{MoraPartiarroyo19a}. In view of the fact that the fitting results are not sensitive to the exact value of the inclination at $\geq 88^\circ$, we simply adopt an inclination angle of $90^\circ$ for the fits. This assumption may lead to an underestimation of $\sigma_{\rm eff}$, which in turn will cause an overestimation of $z_{1}$. Moreover, this effect is more pronounced near the central strip. Nevertheless, even for the central strip, the error introduced by this assumption is still less than 10\%.

\begin{figure*}[th]
    \begin{center}
        \includegraphics[width=\linewidth]{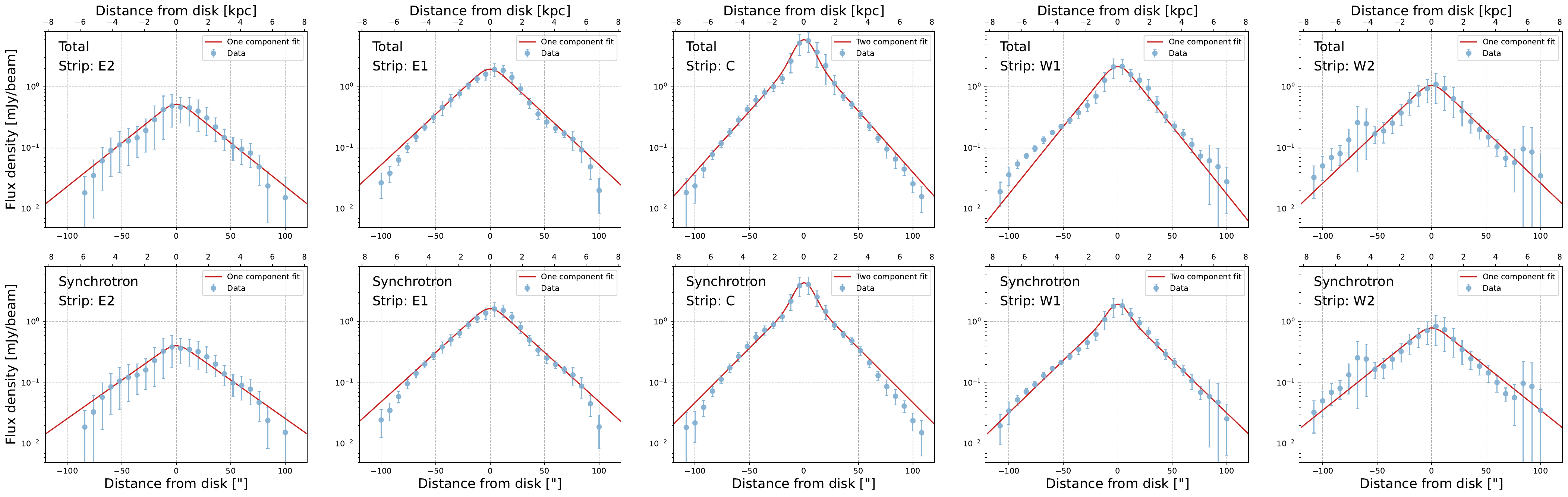}
        \caption{Vertical radio intensity profiles in the five strips of NGC~3556. The top panels display the total intensity data, while the bottom panels display the synchrotron intensity data. The measured data are represented by blue points, and the red lines depict the best-fitting profiles. In cases where a satisfactory fit cannot be achieved using a two-component exponential model, a one-component model is employed as an alternative. }
        \label{fig:vertical_profile}
    \end{center}
\end{figure*}

We present the best-fit profiles of the total intensity map and the synchrotron intensity map in Fig. \ref{fig:vertical_profile}, and list the corresponding fitting parameters in Table \ref{tab:scaleheightfit}. Reasonable results are achievable only in strip C on the total intensity map and in strips C and W1 on the synchrotron intensity map. In the other strips, only one component that represents the halo can be obtained. 

As found in the total intensity map, the scale height of the halo increases from the central strip to the outer ones. Considering that our assumption about the inclination angle $i$ leads to an overestimation of the halo scale height, particularly for the strips closer to the center. \edit1{The actual increase in scale height across the strips may be even more pronounced than our results indicate.} The distribution in the central strip is relatively symmetric, with a slight excess of emission $\sim3$ kpc from the midplane to the south, which corresponds to the bubble-like structure mentioned in \S\ref{subsec:Intensity}. The distribution in strip W1 exhibits local asymmetry, with the northern part consistent with the one-component model and the southern part more consistent with having two components. The situation in Strip E1 is similar to Strip W1, though the northern and southern parts appear to have exchanged. For the outermost strips, E2 and W2, the intensity distributions are in agreement with a one-component exponential model, despite the contamination caused by the scattered point sources. 

By averaging all data extracted from the strips on the total intensity map, we derive a mean scale height of the halo as $z_1 = 1.68\pm 0.29 \mathrm{\ kpc}$. This value is in good agreement with the measurements for other CHANG-ES galaxies \citep{Krause18,MoraPartiarroyo19a,Stein19}. In comparison, the scale height of the halo in NGC~3556 is reported to be $3.3 \pm 0.8 \ \mathrm{kpc}$ in the L-band (1.5 GHz) by \citet{Miskolczi119}. A larger scale height at 1.5 GHz compared to 3 GHz is expected for a synchrotron energy loss-dominated halo \citep{Krause18,Mulcahy18}.

In the synchrotron intensity map, asymmetry is also evident in strips E1 and W1, where we measure the mean scale height of the galactic halo to be $z_1 = 1.93\pm 0.28 \mathrm{\ kpc}$. This larger value than that of the total emission is consistent with the expectation that thermal emission contributes more significantly to the total radio emission in the galactic disk. In addition, the cosmic-ray electrons responsible for synchrotron emission propagate further away from star-forming regions compared to the non-relativistic electrons that generate thermal emission. 

In summary, our analysis of the scale height demonstrates that the exponential function provides a plausible fit to the vertical profile. The resulting scale height of the halo is larger in the synchrotron intensity map than in the total intensity map, indicating the influence of thermal emission and the propagation characteristics of cosmic-ray electrons.

\begin{table*}
    \begin{center}
        \caption{Parameters from scale height fitting of NGC~3556}

        \begin{tabular*}{0.85\linewidth}{lllllll}
            \hline\hline
            Map & Strip & \multicolumn{2}{c}{Disk} & \multicolumn{2}{c}{Halo} & $\chi^{2}_{\mathrm{red}}$\\
            \cline{3-6}
             & & $w_0$ [mJy/beam] & $z_0$ [kpc] & $w_1$ [mJy/beam] & $z_1$ [kpc]& \\
            \hline
            Total & E2 & - & - & $0.61\pm0.03$ & $2.07\pm0.13$& 0.22\\
            & E1 & - & - & $2.34\pm0.11$ & $1.78\pm0.11$ & 1.97 \\
            & C & $7.10\pm1.84$ & $0.27\pm0.19$& $3.82\pm1.46$&$1.48\pm0.33$& 0.88\\
            & W1 & - & - & $2.73\pm0.13$ & $1.33\pm0.09$ & 3.14 \\
            & W2 & - & - & $1.27\pm0.06$ & $1.74\pm0.11$ & 0.46 \\
            \hline 
            Synchrotron& E2 &- &- & $0.46\pm0.02$ & $2.35\pm0.15$ & 0.26\\
            & E1 &- &- &$1.96\pm0.08$ & $1.83\pm0.11$ &1.41 \\
            & C &$5.72\pm1.03$ &$0.24\pm0.08$ &$2.77\pm0.38$ &$1.66\pm0.15$ &1.93 \\
            & W1 &$1.67\pm0.96$ &$0.23\pm0.25$ &$1.63\pm0.30$ &$1.72\pm0.22$ & 0.53\\
            & W2 &- &- &$0.93\pm0.05$ & $2.07\pm0.14$ &0.48 \\
            \hline
        \end{tabular*}
        \label{tab:scaleheightfit}
    \end{center}
%\footnotesize \textbf{Notes:} 
\end{table*}

\subsection{Magnetic field}\label{subsec:MagneticField}

The linear polarization observation allows us to reveal the magnetic field in NGC~3556. %We first calculate the polarized intensity and apparent magnetic field orientations using the Stokes $Q$ and $U$ maps derived from all 14 usable spectral windows (cf. \S\ref{subsec:Polarization}). Subsequently, we apply RM-synthesis to obtain the RM map and the intrinsic magnetic field orientations corrected for Faraday rotation (cf. \S\ref{subsec:RotationMeasure}). We note that the magnetic field strength map is derived on the basis of the equipartition assumption (cf. \S\ref{subsubsec:MagneticFieldStrength}).
\edit1{In this work, we apply the RM synthesis technique (cf. \S\ref{subsubsec:RMsynthesis}) to derive the information about the intrinsic magnetic field, including its components perpendicular and parallel to the line of sight. This technique enables the calculation of the polarized intensity, the intrinsic magnetic field orientations corrected for Faraday rotation (cf. \S\ref{subsec:Polarization}), and the rotation measure (cf. \S\ref{subsec:RotationMeasure}). Additionally, the magnetic field strength map is constructed based on the equipartition assumption (cf. \S\ref{subsubsec:MagneticFieldStrength}).}

\subsubsection{Polarization} \label{subsec:Polarization}

In \edit1{the top panel of Fig. \ref{fig:RM_map}}, the polarized intensity contours and \edit1{intrinsic} magnetic field orientations in the S-band are overlaid on the total intensity map. The polarized emission exhibits a clumpy distribution across the entire galaxy. The total polarized flux intensity, enclosed by \edit1{the $5\sigma$ polarization contour, is $8.3\pm 0.3 \mathrm{\ mJy}$}. 

In some CHANG-ES galaxies, large-scale X-shaped magnetic field patterns are observed, where the magnetic field is parallel to the disk in the midplane, though transitions to be perpendicular in the halo (e.g., NGC 4217, \citealt{Stein20}; NGC 4631, \citealt{MoraPartiarroyo19b}; NGC 4666, \citealt{Stein19_4666}). In contrast, the magnetic field in NGC~3556 is not globally well-organized, but displays a more complex structure featured by numerous small patches with varying local orientations. The peak of the polarized intensity is located towards the south of the disk, corresponding to the position of the above mentioned bubble-like structure discussed in \S\ref{subsec:Intensity}. In comparison to the bubble-like structure seen in the total intensity, a more spatially extended C-shaped structure is seen in the polarized intensity, where the magnetic field is aligned with the extension. 
In addition, we observe an extended structure above the disk, a feature potentially related to the shell-like structure spotted in the total intensity map. These extensions above and below the central disk may hint at an interaction between the disk and halo \citep{Heald09,Heesen11}, on which further discussion is given in \S\ref{subsec:BField}.

% \begin{figure}[th]
%     \begin{center}
%         \includegraphics[width=8.5cm]{./mag_field_vector.pdf}
%         \caption{Total intensity image of NGC~3556 at S-band with robust zero weighting and a \texttt{'10 arcsec'} uv-taper (same as Fig. \ref{fig:total_intensity}, panel(b)). Green contours at 3 and $270\sigma$ levels with a $\sigma$ of 5.5 $\mathrm{mJy\ beam}^{-1}$ highlight the edge and the central disk respectively. Gray polarization contours are displayed at 3, 6, and 12$\sigma$ levels with the same $\sigma$ value. The apparent magnetic field orientations are represented in blue and clipped below $3\sigma$. The beam size of $10.47''\times 10.15''$ is displayed as a blue ellipse in the bottom-left corner.}
%         \label{fig:pol_map}
%     \end{center}
% \end{figure}

\subsubsection{Rotation Measure} \label{subsec:RotationMeasure}

\begin{figure*}[th]
    \begin{center}
        \includegraphics[width=8.5cm]{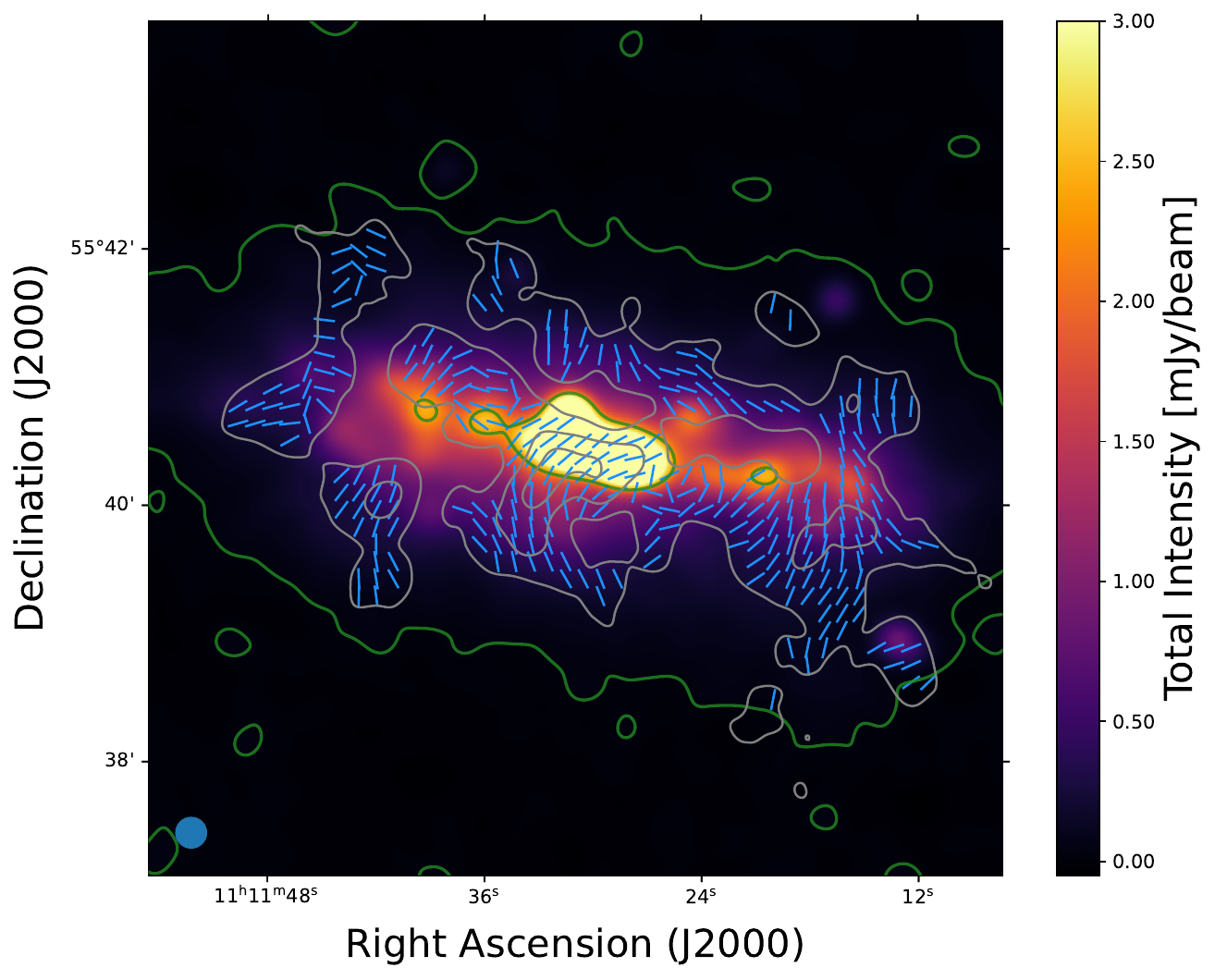}\\
        \includegraphics[width=8.5cm]{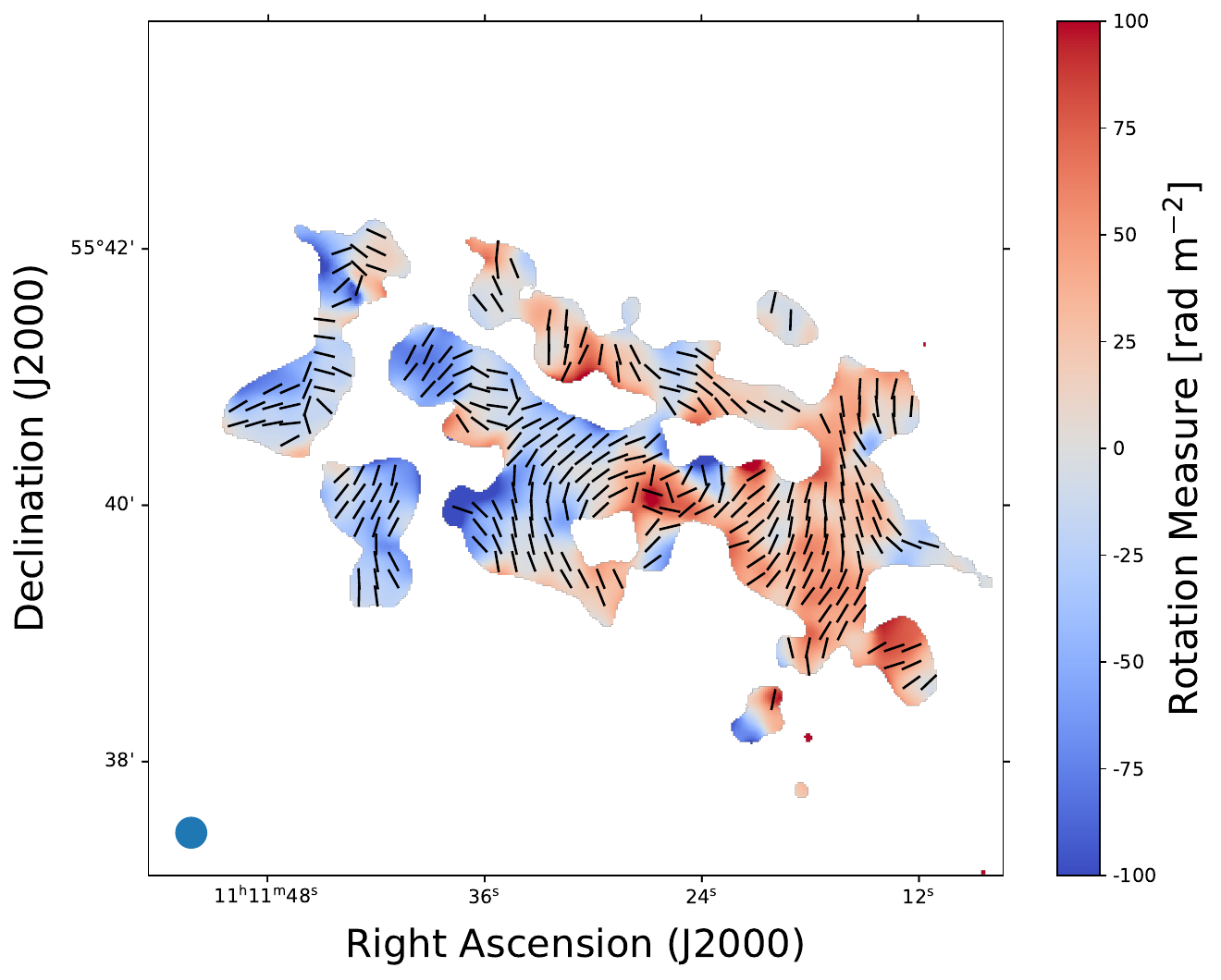}
        \includegraphics[width=8.5cm]{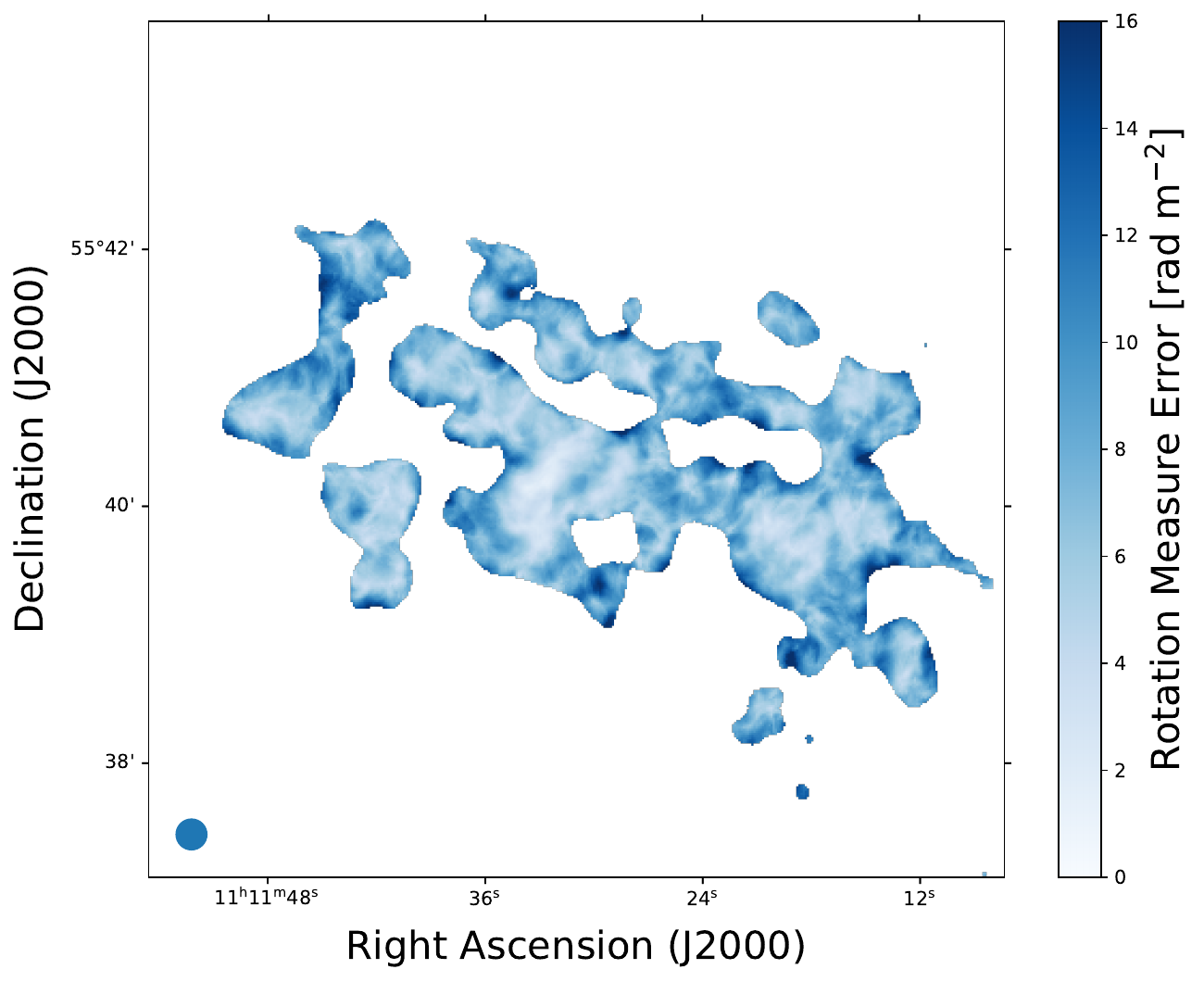}
        \caption{NGC~3556 images from RM-synthesis. \textit{Top}: total intensity color image smoothed to $15''$ with gray polarization contours at 5, 10, and 15$\sigma$ levels with $\sigma$ of 9.71 $\mu\mathrm{Jy\ beam}^{-1}$. Green contours of total intensity at 3 and $270\sigma$ levels with a $\sigma$ of 8.5 $\mu\mathrm{Jy\ beam}^{-1}$ highlight the edge and the central disk, respectively.   \textit{Bottom}: RM map and RM error map of NGC~3556\edit1{, with Galactic foreground subtraction applied,} cut to a level of $5\sigma$ of the polarization intensity map. The intrinsic magnetic field orientations corrected for Faraday rotation, are clipped below $5\sigma$ and displayed in blue and black respectively. The beam size of $15''$ is given in the bottom-left corner.}
        \label{fig:RM_map}
    \end{center}
\end{figure*}

%The RM synthesis technique (cf. \S\ref{subsubsec:RMsynthesis}) enables us to deduce the intrinsic magnetic field vectors for both the components perpendicular and parallel to the line of sight. The results are shown in Fig.~\ref{fig:RM_map}. In the top panel, we overlay the contours of the polarized emission obtained through RM-synthesis on the total intensity map shown in Fig.~\ref{fig:total_intensity}b. Comparing this RM synthesis corrected map with the original polarization map (Fig.~\ref{fig:pol_map}), we find their morphology to be similar, despite a number of extended features in the western disk and the northeastern halo. Furthermore, the peak of the polarized intensity is also located on the southern side of the disk, stretching into the halo, and is aligned with the bubble-like structure seen in the total intensity.

The RM map, corrected for the Galactic foreground, is shown in the bottom panel of Fig. \ref{fig:RM_map}. The RM in NGC~3556 roughly ranges from $-100$ to $100$ $\mathrm{rad\ m}^{-2}$ with an average uncertainty of $8 \mathrm{\ rad\ m}^{-2}$ \edit1{(error of ${\rm RM}_{\rm fg}$ has been included)}, where positive/negative values correspond to a magnetic field pointing toward/away from the observer. Using the observed \ion{H}{1} velocities \citep{Zheng22}, we conclude that the eastern side of the galaxy is approaching us. Analysis of the RM map reveals that predominantly negative/positive RM values appear on the approaching/receding side, suggestive of an overall axisymmetric magnetic field pattern. In the central region, the magnetic field in the southern part points away from/toward the observer on the eastern/western side, which is consistent with the overall trend; while the northern part exhibits an opposite trend, with the eastern/western side pointing towards/away from the observer, hinting at an antisymmetric magnetic field structure in the central region. 

%We deduce the intrinsic magnetic field orientations by applying the Faraday rotation correction. For $|\mathrm{RM}|\lesssim 100 \mathrm{\ rad\ m}^{-2}$, the rotation at S-band should be $\lesssim 60^{\circ}$. This becomes discernible when the corrected maps are compared to the uncorrected ones, where we find similarities in most of the regions, with differences noticeable only in localized patches. 

\subsubsection{Magnetic field strength} \label{subsubsec:MagneticFieldStrength}

The magnetic field strength in NGC~3556 is derived based on the equipartition assumption, which postulates energy equilibrium between the total cosmic rays and the magnetic field. This assumption provides a convenient method to infer magnetic field information from synchrotron intensity and is valid globally in normal star-forming galaxies and locally at scales larger than $\sim 1 {\rm\ kpc}$ \citep{Seta19}. We derive the magnetic field strength map using the revised equipartition formula presented in \cite{Beck05}:

\begin{equation} \label{equ:magnetic_field}
    B_{\mathrm{eq}}=\left[\frac{4 \pi(1-2\alpha )(K_{0}+1)I_{\mathrm{nt}}E_{\mathrm{p}}^{1+2\alpha}(\nu/2c_{1})^{-\alpha}}{(-2\alpha -1 )c_{2}(-\alpha)l c_{4}}\right]^{\frac{1}{3-\alpha}}.
\end{equation}

Here, $\alpha$ represents the spectral index of the nonthermal radio emission defined as $I_{\nu}\propto\nu^{\alpha}$, where $I_{\nu}$ is the nonthermal intensity at frequency $\nu$.
$K_{0}$ is the number density ratio of protons to electrons and is assumed to be $K_0 = 100$ in this work. $E_{\mathrm{p}}$ is the proton rest energy and $c_{1-4}$ are constants. The path length along the line of sight through the galaxy is denoted as $l$, and a typical value of 20 kpc is assumed for the halo region of NGC~3556. It is worth noting that the estimation of the magnetic field strength in Eq. \ref{equ:magnetic_field} exhibits a weak dependence on the path length, with $B_{\mathrm{eq}}\propto l^{-1/(3-\alpha)}$. Therefore, assuming a constant path length minimally affects our results.

\begin{figure}[th]
    \begin{center}
        \includegraphics[width=8.5cm]{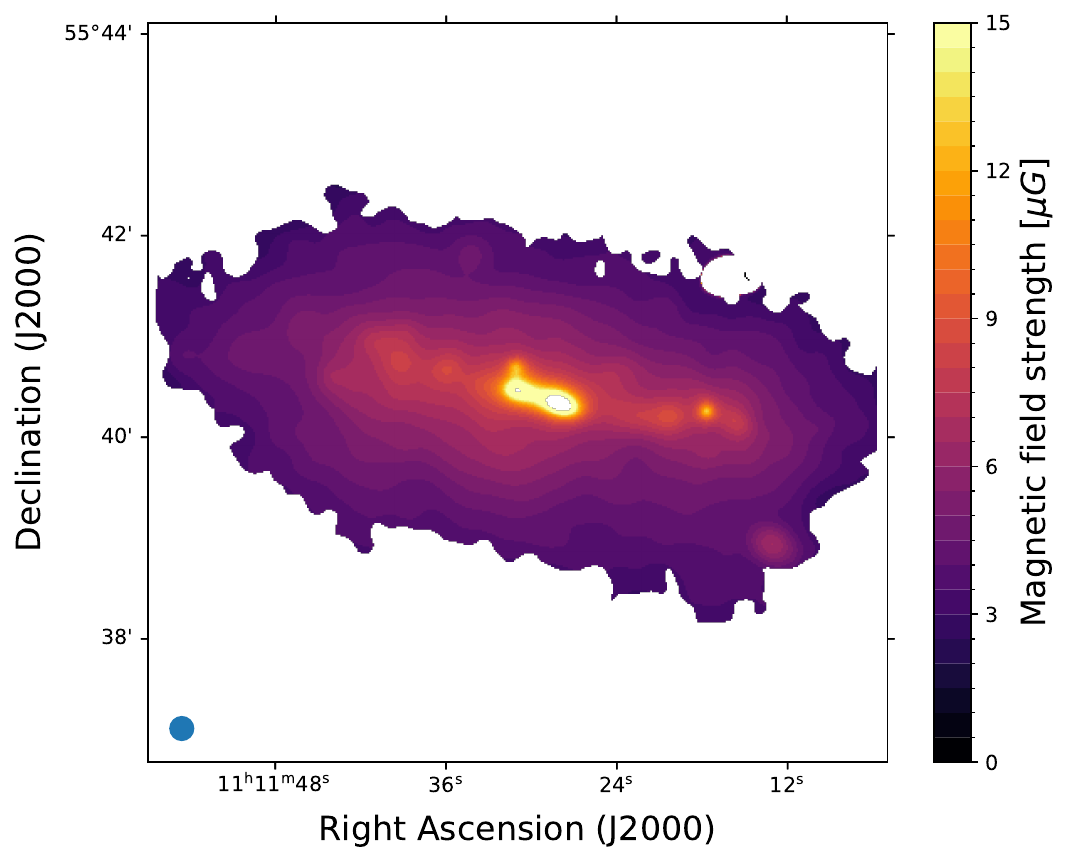}
        \caption{Magnetic field strength map of NGC~3556 calculated using the revised equilibrium formula. Some pixels are excluded due to spectral indices exceeding $-0.5$. The blue circle in the bottom-left corner represents a beam size of $15'' \times 15''$.}
        \label{fig:magnetic_field_strength}
    \end{center}
\end{figure}

Using the nonthermal intensity and the nonthermal spectral index maps (cf. \S\ref{subsec:ThermalNonthermalMap}) and employing the revised equilibrium formula, we perform a pixel-by-pixel calculation of the magnetic strength, and present the result in Fig. \ref{fig:magnetic_field_strength}.
We find the average magnetic field to be $8.3\ \mu\mathrm{G}$ in the disk, $13\ \mu\mathrm{G}$ in the center, and $4.6\ \mu\mathrm{G}$ in the southern halo and $4.4\ \mu\mathrm{G}$ in the northern halo. Assuming a 50\% uncertainty in both $K_{0}$ and $l$, we estimate an error of $\lesssim2\ \mu {\rm G}$ in the disk and $\lesssim0.7\ \mu{\rm G}$ in the halo. These values are consistent with the magnetic field strength reported for NGC~3556 in \cite{Miskolczi119}, where a 144 MHz map and a spectral index map (144 MHz vs. 1.5 GHz) are used, although no thermal decomposition is performed.
The magnetic field in NGC~3556 is relatively weaker than in 13 other CHANG-ES galaxies \citep[9--11 $\mu\mathrm{G}$ over the total galaxy and 11--15 $\mu\mathrm{G}$ in the galactic disk;][]{Krause18}. 

\subsection{Point sources} \label{subsec:PointSources}

\begin{figure*}[th]
    \begin{center}
        \includegraphics[width=\linewidth]{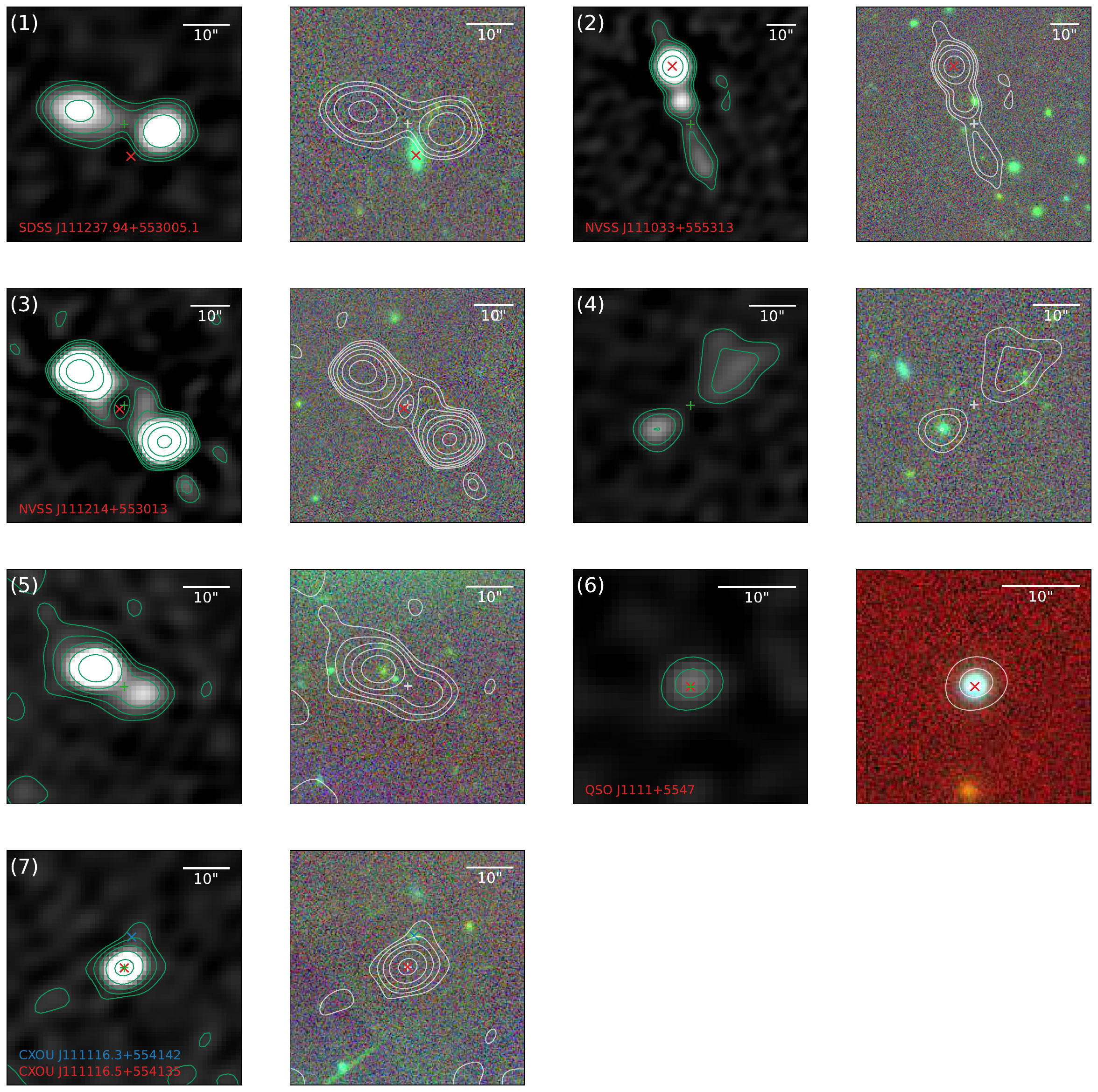}
        \caption{Cutouts of the point sources surrounding NGC~3556. For each source, the left panel shows the S-band total intensity image, while the right panel presents the tri-color image made from the Pan-STARRS \textit{y, i, g} filters. Contours start at $3\sigma$ above the background noise and increase by a factor of 2. The center of each source is marked by a plus sign. The scale bar in the upper right corner indicates the size of each image. Cross-matched sources are marked with blue and red crosses. Relevant parameters are provided in Table \ref{tab:point_source}.}
        \label{fig:point_source}
    \end{center}
\end{figure*}

\begin{table*}
    \caption{Parameters of the detected point sources}
        \begin{center}
            \begin{tabular*}{\linewidth}{ccccccc}
                \hline\hline
                ID& R.A. (J2000)& Decl. (J2000) & Flux density & Cross-matched Sources & Redshift& References \\
                & & & (mJy) & &\\
                \hline
                1 & 11:12:38.12 & +55:30:12.01 & $ 0.92 \pm 0.06 $ & SDSS J111237.94+553005.1 &0.13754& \cite{Alam15}\\
                2 & 11:10:32.59 & +55:52:58.69 & $ 1.49 \pm 0.10 $ & NVSS J111033+555313 && \cite{Mahatma19}\\
                3 & 11:12:14.16 & +55:30:14.95 & $ 4.22 \pm 0.11 $ & NVSS J111214+553013 && \cite{Condon98}\\
                4 & 11:10:29.39 & +55:45:15.93 & $ 0.27 \pm 0.04 $ & && \\
                5 & 11:11:12.41 & +55:38:53.01 & $ 1.09 \pm 0.06 $ & && \\
                6 & 11:11:32.19 & +55:47:26.08 & $ 0.04 \pm 0.01 $ & QSO J1111+5547 &0.76633& \cite{Allen11} \\
                7 & 11:11:16.55 & +55:41:35.82 & $ 0.49 \pm 0.03 $ & {\begin{tabular}{c} CXOU J111116.5+554135 \\ CXOU J111116.3+554142 \\ \end{tabular}} && \cite{Wang03}\\
                \hline
            \end{tabular*}
            \label{tab:point_source}
        \end{center}
    % \footnotesize \textbf{Notes:} 
\end{table*}

Fig. \ref{fig:point_source} displays cutouts of the point sources detected in the field of view, whose parameters are listed in Table \ref{tab:point_source}. Sources 1--5 exhibit double-lobe morphology, among which sources 1--3 can \edit1{each} be cross-matched with a galaxy. Source 6 is a UV-bright background AGN at \edit1{a redshift of} $ z \approx 0.77 $, characterized by multiple UV absorption lines detected by \cite{Li24}. Source 7, already discussed in \S\ref{subsec:ThermalNonthermalMap}, is spatially associated with two X-ray sources: one coinciding with its center and the other offset by $ 6.8'' $. The coincidence of this source with the X-ray, also seen in L and C-band, was first pointed out in \cite{irw22}.

\section{Discussion} \label{sec:Discussions}

\subsection{Vertical radio intensity profiles and constraint on CR-driven wind} \label{subsec:VerticalProfileWind}

\begin{figure*}[th]
    \begin{center}
        \includegraphics[width=0.47\linewidth]{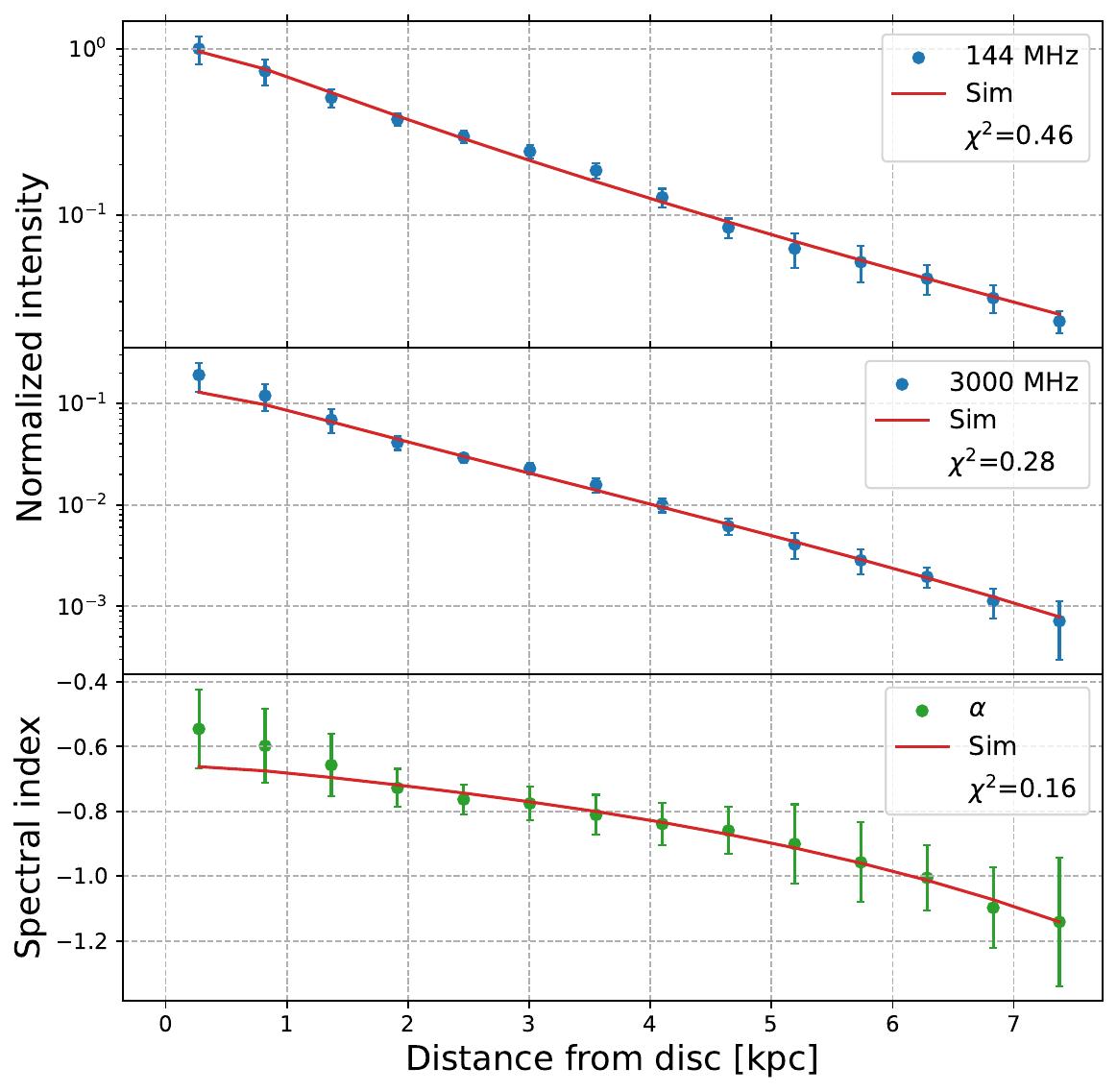} \qquad
        \includegraphics[width=0.47\linewidth]{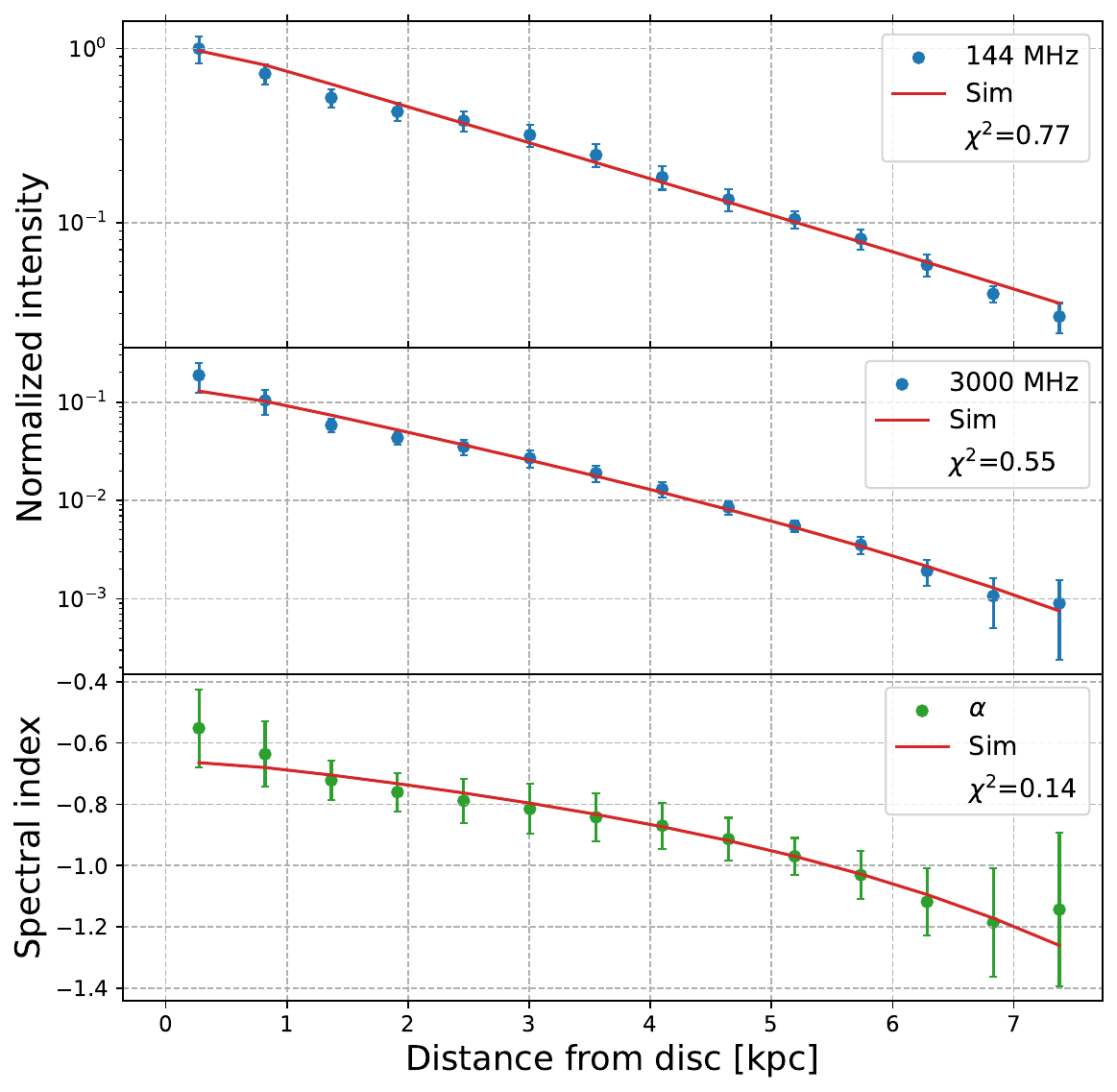}
        \caption{Cosmic ray advection transport models for NGC~3556 using \texttt{SPINNAKER}. \textit{Left panel}: Northern segment of the central strip. \textit{Right panel}: southern segment of the central strip. Each panel depicts \texttt{SPINNAKER} fitting of the LOFAR 144 MHz (top graph) and VLA 3.0 GHz (middle graph) intensity profiles, as well as the spectral index profile (bottom graph), simultaneously. The red lines represent the model simulations. The relevant parameters are listed in Table \ref{tab:CR_models}.}
        \label{fig:CR_models_advection}
    \end{center}
\end{figure*}

\begin{figure*}[th]
    \begin{center}
        \includegraphics[width=0.47\linewidth]{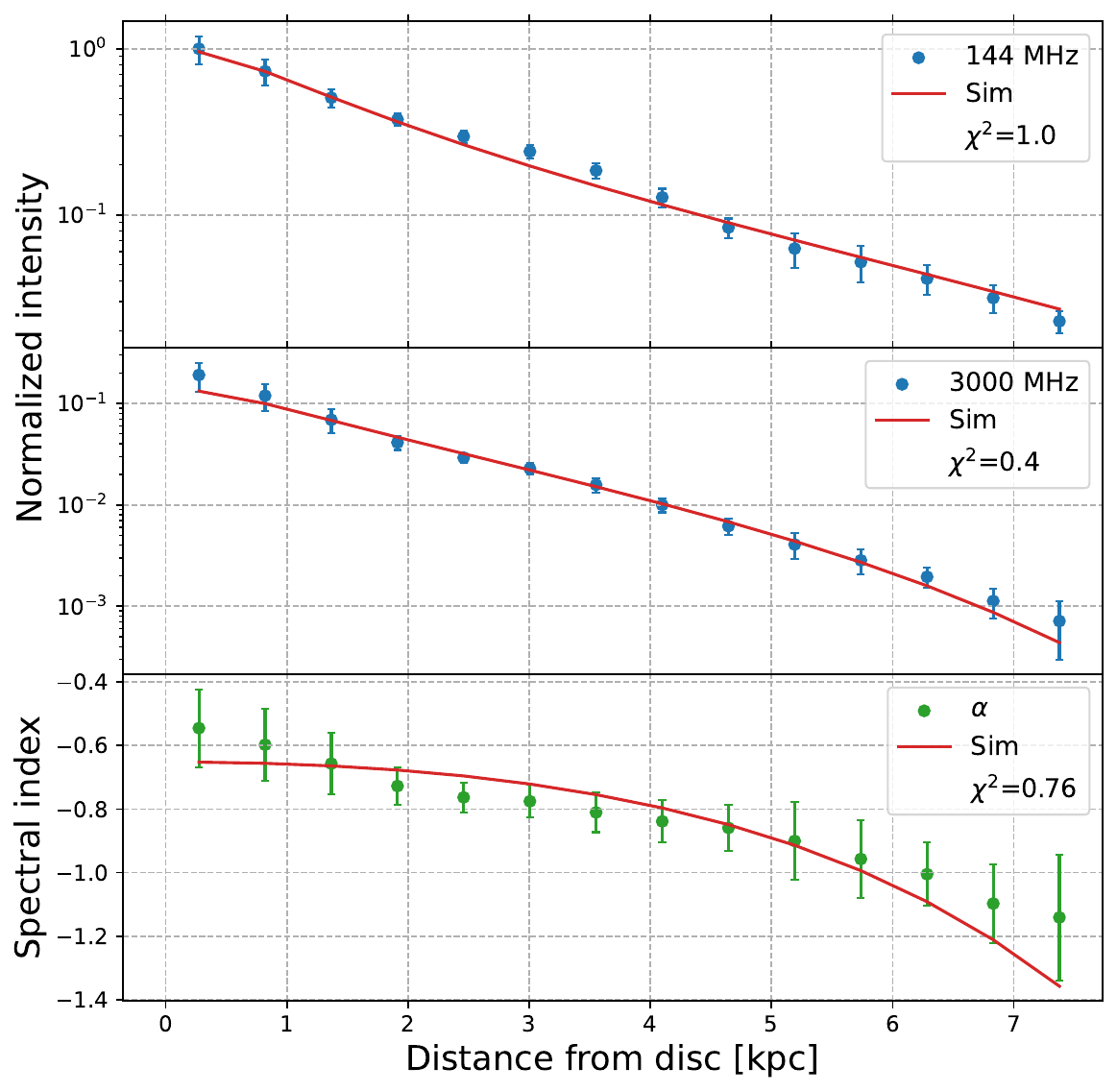} \qquad
        \includegraphics[width=0.47\linewidth]{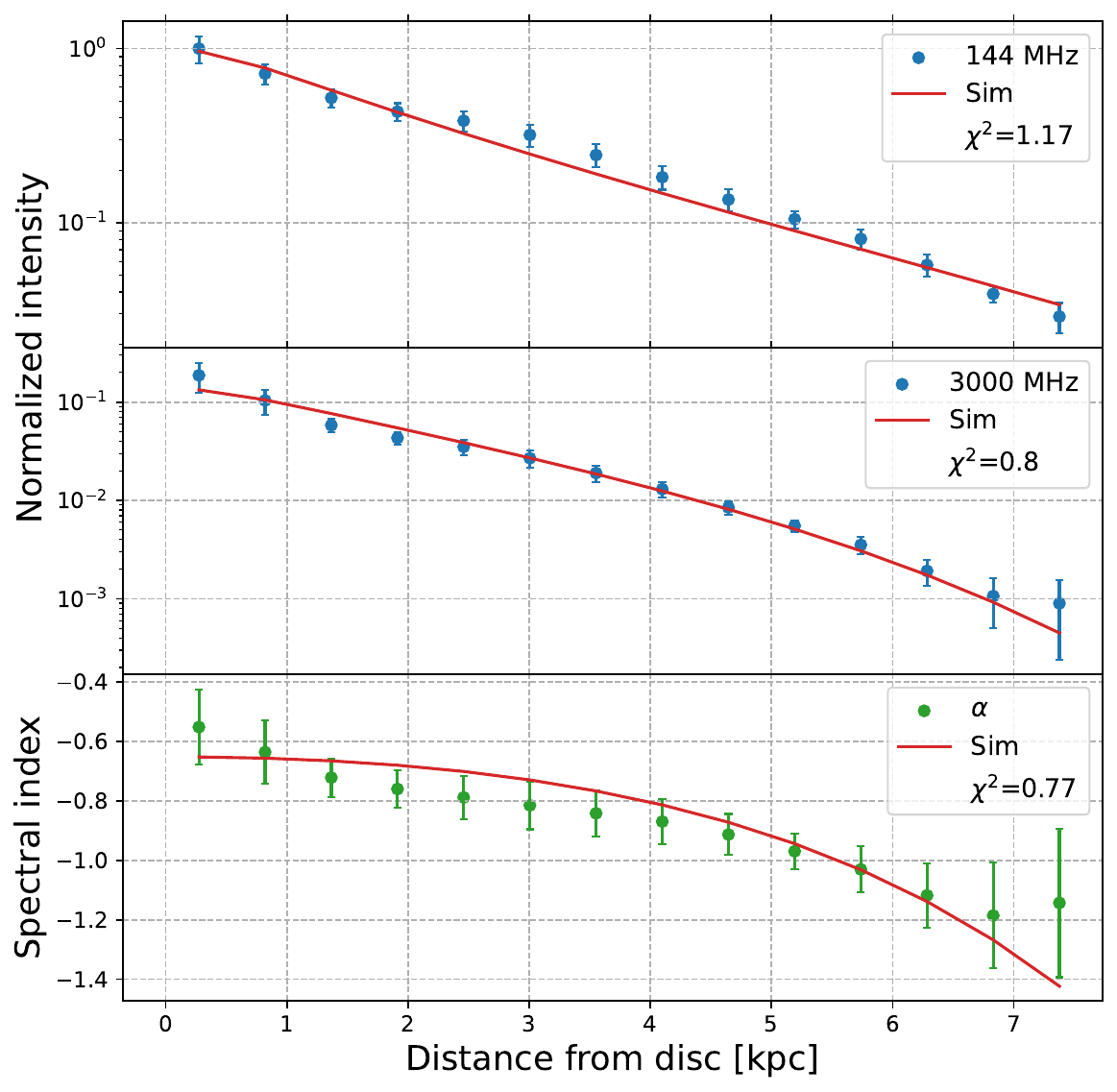}
        \caption{Cosmic ray diffusion transport models for NGC~3556 using \texttt{SPINNAKER}. \textit{Left panel}: Northern segment of the central strip. \textit{Right panel}: southern segment of the central strip.  The relevant parameters are listed in Table \ref{tab:CR_models}.}
        \label{fig:CR_models_diffusion}
    \end{center}
\end{figure*}

\begin{table*}
\caption{Best-fit parameters of the cosmic-rays transport models}
    \begin{center}
        \begin{tabular*}{0.76\linewidth}{lcccccccccc}
            \hline\hline
            Parameter & \multicolumn{5}{c}{Northern Segments}& \multicolumn{5}{c}{Southern Segments}\\
            \cmidrule(lr){2-6}\cmidrule(lr){7-11}
            & E2 & E1 & C & W1 & W2 &E2 & E1 & C & W1 & W2 \\
            \hline
            $\gamma$&2.3&-&-&-&-&-&-&-&-&-\\
            $B_{0}\ (\mu \mathrm{G})$ &10&-&-&-&-&-&-&-&-&-\\
            $B_{1}\ (\mu \mathrm{G})$ &6&-&-&-&-&-&-&-&-&-\\
            \hline
            \multicolumn{11}{c}{Diffusion model}\\
            \hline
            $h_{B1}\ (\mathrm{kpc})$
            &3.5&2.5&1.5&2.0&3.5&3.0&3.0&2.0&0.5&3.0\\
            $h_{B2}\ (\mathrm{kpc})$
            &13.0&8.0&6.5&7.0&4.0&47.5&11.5&7.0&10.5&32.0\\
            $D_{0}\ (10^{28}\ \mathrm{cm}^{2}\mathrm{s}^{-1})$
            &12&11&7&13&15&8&8&9&3&17\\
            $\mu$
            &0.0&0.1&0.2&0.0&0.2&0.0&0.1&0.1&0.4&0.0\\
            $\chi^{2}_{\mathrm{red}}$
            &0.14&0.40&0.72&0.16&0.22&0.19&0.44&0.91&0.92&0.28\\
            \hline
            \multicolumn{11}{c}{Advection model (constant speed)}\\
            \hline
            $h_{B1}\ (\mathrm{kpc})$
            &5.5&4.0&2.0&2.5&3.5&4.5&4.5&3.5&2.0&4.5\\
            $h_{B2}\ (\mathrm{kpc})$
            &5.5&4.0&5.0&5.5&4.0&12.0&5.0&4.0&8.5&13.5\\
            $V_{0}\ (\mathrm{km}\mathrm{\ s}^{-1})$
            &250&260&230&240&410&170&210&210&230&260\\
            $\chi^{2}_{\mathrm{red}}$
            &0.11&0.35&0.30&0.07&0.20&0.09&0.12&0.49&0.55&0.16\\
            \hline
        \end{tabular*}
        \label{tab:CR_models}
    \end{center}
\footnotesize \textbf{Notes:} $\gamma$ (fixed), power law index of the cosmic ray injection spectrum; $B_{0}$ (fixed), magnetic field strength in the galactic midplane; $B_{1}$ (fixed), magnetic field strength of the thin disk component; $h_{B1}$, scale height of the magnetic field of the thin disk component; $h_{B2}$, scale height of the magnetic field of the thick disk component; $D_{0}$, diffusion coefficient at a CRE energy of 1 GeV; $\mu$, energy dependence of the diffusion coefficient with $D=D_{0}(E/1\ \mathrm{GeV})^{\mu}$; $V_{0}$, wind speed; $\chi^{2}_{\mathrm{red}}$, reduced chi square of the best-fitting model computed as the arithmetic mean of the two intensity profiles and the spectral index profile.
\end{table*}

The scale height analysis in \S\ref{subsec:ScaleHeight} shows that exponential functions can plausibly describe the vertical radio intensity profiles in S-band. Due to the complication of the fitting function, we obtain acceptable results only in strip C using a two-component exponential fitting. As for the other strips, one-component exponential function is superior to a Gaussian function. This suggests that advection may be the dominant cosmic-ray transport mechanism in NGC~3556 \citep{Heesen16}. 

We further investigate cosmic ray transport using the 1D CR transport model, SPectral INdex Numerical Analysis of K(c)osmic-ray Electron Radio-emission \citep[SPINNAKER,][]{Heesen16}. This model is applied to the total intensity profiles of the synchrotron maps, which are divided into 5 strips, each with a box size of $70''\times 8''$ for the S-band and LOFAR data, similar to the settings in \S\ref{subsec:ScaleHeight}. Due to asymmetry of the intensity distribution, we further divide each strip into northern and southern segments, resulting in a total of ten profiles for analyzing the cosmic ray propagation process. 

The models assume either pure diffusion, characterized by a diffusion coefficient $D=D_{0}(E/1\ \mathrm{GeV})^{\mu}$, or advection with a constant advection speed $V_{0}$. Other critical, adjustable parameters include the CRe injection index $\gamma$ and the magnetic field parameters. The injection index relates to the nonthermal spectral index via $\alpha_{\mathrm{nth}}=(1-\gamma)/2$. The magnetic field is modeled as a double exponential function with contributions from both a thin and a thick disk:

\begin{equation}
    B(z) = B_{1}\exp (-z/h_{B1})+(B_{0}-B_{1})\exp(-z/h_{B2}),
\end{equation}
where $B_{0}$ is the magnetic field strength in the disk, $B_{1}$ is the magnetic field strength of the thin disk component, and $h_{B1}$ and $h_{B2}$ are the scale heights of the thin and thick disks, respectively.

To achieve the best-fit results, we minimize the reduced chi-square ($\chi^{2}_{\mathrm{red}}$) value, which is defined as the average of the reduced chi-square values resulting from fits of the intensity and of the spectral index profiles. Several parameters are fixed based on the spectral index map and equilibrium magnetic field map obtained in \S\ref{subsec:ThermalNonthermalMap} and \S\ref{subsec:MagneticField}, in particular, $\gamma=2.3$, $B_{0}=10\ \mu \mathrm{G}$, and $B_{1}=6\ \mu \mathrm{G}$. 

Fig. \ref{fig:CR_models_advection} and \ref{fig:CR_models_diffusion} display the best-fit profiles of the central strip for advection and diffusion, respectively, with the corresponding parameters for all strips presented in Table \ref{tab:CR_models}. Both advection and diffusion models fit the data reasonably, with $\chi^{2}_{\mathrm{red}}$ values for each strip less than one. \edit1{This is likely due to overfitting, as the limited dataset (VLA S-band and LOFAR 144 MHz) relative to the number of model parameters reduces the ability of the models to impose strong constraints on the data.} Fits to advection model yield slightly favorable results, as evidenced by lower $\chi^{2}_{\mathrm{red}}$ values across each strip. The primary difference in these fits arises from the spectral index profiles: diffusion results in a parabolic shape, whereas advection produces a linear profile. The advection model matches our data with insignificant advantage, though the difference is visually discernible. \edit1{Future works incorporating additional bands, such as C-band and L-band, or higher-resolution data from instruments like the SKA, could further constrain the model parameters and improve their reliability.}

When an advection model is adopted, the best-fit advection speeds are consistent across all strips with the sole exception of strip W2. This is likely due to the point sources in strip W2, located at $z\sim 6.0\mathrm{\ kpc}$ in the northern segment, and $z\sim 4.5\mathrm{\ kpc}$ in the southern segment. 
Excluding stripe W2, we find the mean advection speeds in the northern segment to be slightly higher ($245 \pm 15 \mathrm{\ km}\mathrm{\ s}^{-1}$) than in the southern segment ($205 \pm 25 \mathrm{\ km}\mathrm{\ s}^{-1}$). We compare these speeds to the escape velocity of the galaxy following \citet{Veilleux05} and \citet{Miskolczi119}:

\begin{equation}
    v_{\rm esc}(r)= \sqrt{2}v_{\rm rot}\sqrt{1+\ln{(R_{\rm max}/r)}},
\end{equation}
where $r_{\rm rot}$ is the rotation velocity, which is $\sim 150 {\rm \ km\ s}^{-1} $ for NGC~3556 \citep{Li24}, and $R_{\rm max}$ is the outer radius of the truncated isothermal sphere, with $r$ being the radius in spherical coordinates. We obtain a range of $v_{\mathrm{esc}}\approx 190-300\mathrm{\ km}\mathrm{\ s}^{-1}$ at the limiting outermost region of the observable halo ($z\sim 7\mathrm{\ kpc}$), when we vary $R_{\rm max}$ between 10 and 20 kpc and the radial distance in the galactic plane between 0 and 10 kpc. We find that the advection speeds are comparable to the range of the escape velocity. This is consistent with the correlation found in 11 edge-on galaxies by \citet{Heesen18} and can be explained by a cosmic ray-driven wind. 

In addition, to assess the possibility of CRe escaping the galaxy, we calculate the synchrotron lifetime using the equation from \cite{Krause18}:

\begin{equation}
    t_{\mathrm{syn}}[\mathrm{yr}]=1.06\times10^{9}
    \left(\frac{\nu}{\mathrm{GHz}}\right)^{-\frac{1}{2}}\left(\frac{B}{\mu\mathrm{G}}\right)^{-\frac{3}{2}}.
\end{equation}

Adopting an S-band frequency of 3 GHz and an average magnetic field strength of 6 $\mu\mathrm{G}$, the magnetic field strength map shown in Fig. \ref{fig:magnetic_field_strength} renders a synchrotron lifetime of approximately 40 Myr. Therefore, the velocity required for CRe to reach the halo's outer limit of 7 kpc is about $170 \mathrm{\ km}\mathrm{\ s}^{-1}$. This is lower than the velocities obtained from fitting the data to the advection model, indicating that CRe in NGC~3556 may indeed reach the observable outskirts before their energy is exhausted by synchrotron emission. 

Therefore, we conclude that in NGC~3556, cosmic ray transport is primarily dominated by advection. This conclusion is supported by the exponential intensity profile and the linear spectral index profile \citep{Heesen16}. For simplicity, we assume a constant advection speed in the 1D CR transport model analysis. The results indicate that CRs are efficient in transportation to the outer edge of the galaxy, rendering the halo as observed in the S-band. 

\subsection{Topological structure of the magnetic field} \label{subsec:BField}

The linear polarization data of NGC~3556, combined with RM-synthesis, allow us to elucidate the magnetic field structures in the galactic disk and halo.

\edit1{The} top panel of Fig. \ref{fig:RM_map} demonstrate\edit1{s} that polarized emission is present over the entire disk, except for a small region on the eastern side, where the oddness may be attributed to Faraday depolarization. In contrast to the large-scale X-shaped magnetic field patterns observed in some CHANG-ES galaxies \citep{MoraPartiarroyo19b,Stein20,Krause20}, which is interpreted as a quadrupolar field in the halo \citep{Braun10}, NGC~3556 exhibits small-scale magnetic field structure with numerous local patterns displaying varying local orientations, depicting a spatially variable configuration.

%When comparing the distribution of polarized intensity between the approaching (eastern) and receding (western) sides of NGC 3556, we observe that approximately 55\% of the polarized emission is concentrated on the approaching side, with the remaining 45\% on the receding side. This asymmetry may result from stronger Faraday depolarization on the receding side. A similar pattern is observed in the L-band citep{Miskolczi119}, where polarized emission is only detected on the approaching side due to the more pronounced Faraday effect at lower frequencies. However, the trend reverses in the C-band, with slightly higher polarization on the receding side citep{Krause20}. In addition, citet{Krause20} also found that in 13 out of 18 CHANG-ES galaxies, stronger polarization is observed on the approaching side in the C-band.

The strongest polarized intensity in NGC~3556 is located on the southern side of the disk, accompanied by an arc-shaped polarized intensity structure extending several kpc toward the southern halo. The magnetic field orientation is aligned with this structure, indicating the existence of a large-scale magnetic field line loop. Panels (c) and (d) of Fig. \ref{fig:total_intensity} suggest that this polarization feature spatially coincides with a bubble-like structure in the total intensity map, where ionized gas is clearly traced by H$\alpha$ emission. Additionally, numerous \ion{H}{1} extensions and superbubbles have been observed emerging from the galactic plane, as reported by \citet{king97}. A corresponding shell-like structure is found on the northern side, with an extended structure seen in polarization seemingly associated with it. These observations suggest the possibility of a bipolar bubble-like structure stretching from the central disk. Such a bipolar structure is also detectable in other galaxies \citep{Li19,Stein20} and reproducible in simulations \citep{Pillepich21,Pfrommer22}. A possible explanation is wind-blown superbubbles driven by the pressure from both cosmic rays and ionized gas resulting from supernovae activity in the central disk.

Compared to the RM map generated at C-band \citep[see Fig. A.8 in][]{Krause20}, which is primarily confined to the galactic disk, the RM map obtained at S-band provides a more comprehensive view that extends into the halo. The RM map presented in the bottom panel of Fig. \ref{fig:RM_map} indicates that the overall magnetic field may exhibit a toroidal configuration, similar to the results of \citet{Krause20}. In particular, the magnetic field points away from the observer on the eastern side and towards the observer on the western side, opposite to the direction of galactic rotation. Simulations by \citet{Henriksen21} demonstrate that this global toroidal configuration can be reproduced by a dipolar field, taking into account the effects of turbulent dynamo and halo rotational lag. In the northern part of the central region, the RM signs appear to be reversed, which may imply the presence of a reversed field above and below the galactic plane \citep{Myserlis21}, or could be influenced by the central spiral arms. However, given the limited spatial resolution and the significant uncertainties associated with the RM measurements, our understanding of the magnetic field structure above the disk remains tentative.

\section{Summary and Conclusions} \label{sec:Summary}

In this paper, we present the VLA radio continuum observation of the nearby late-type spiral galaxy NGC~3556, a member of the CHANG-ES survey sample, in S-band (3 GHz) in the C array configuration. The high inclination of NGC~3556 allows for an in-depth probe of CR propagation and magnetic field structures in the galactic halos. We analyze CR propagation using 1D cosmic-ray transport models and vertical scale height fits. In addition, RM-synthesis is applied to reveal the magnetic field structure. Our main results are summarized as follows:

\begin{itemize}
    \item[-] The total radio intensity distribution of NGC~3556 in S-band exhibits a box-like shape. The detectable halo extends approximately 6.7 kpc from the galactic plane. The measured total flux density in S-band is $155 \pm 5 \mathrm{\ mJy}$. 
    \item[-] We decompose the thermal and synchrotron components of the total radio emission by estimating the thermal contribution based on H$\alpha$ and 24$\mu$m data. The average thermal fraction is $\sim20\%$ within the disk, decreasing to $\sim 5\%$ in the halo.
    \item[-] Radio scale heights at various radii relative to the major axis are determined by fitting an exponential profile to the vertical distributions of both total and synchrotron intensity. The average radio scale height of the halo is $1.68\pm 0.29 \mathrm{\ kpc}$ for total intensity and is $1.93\pm 0.28 \mathrm{\ kpc}$ for non-thermal intensity, as a result of the longer lifetime of the cosmic-ray electrons responsible for synchrotron emission.
    \item[-] Using additional LOFAR observation at 144 MHz, we estimate the nonthermal spectral indices and the equipartition magnetic field strength. The nonthermal spectral index is, on average, $\sim-0.64$ in the disk and $\sim-0.85$ in the halo. The equipartition magnetic field strength is estimated to be $\sim8.3\ \mu\mathrm{G}$ in the disk and $\sim4.5\ \mu\mathrm{G}$ in the halo, which are insignificantly lower than the values reported for other CHANG-ES galaxies \citep{Krause18}.
    \item[-] Applying 1D cosmic-ray transport models, we find that the advection model fits the data more reasonably than the diffusion model. Our best-fit model renders advection speeds of $245 \pm 15 \mathrm{\ km}\mathrm{\ s}^{-1}$ and $205 \pm 25 \mathrm{\ km}\mathrm{\ s}^{-1}$ for the northern and southern segments, respectively. These values are in line with the estimation of the velocity required for the CRe to reach the halo's outer edge.
    \item[-] The linearly polarized emission in S-band is detected patchily across NGC~3556. Unlike the large-scale X-shaped magnetic field patterns observed in some other CHANG-ES galaxies, NGC~3556 exhibits various small patches with varying local orientations. This complex structure indicates a less uniform and more turbulent magnetic field environment, rather than a clearly defined quadrupolar field.
    \item[-] Results from RM-synthesis indicate that the magnetic field points away from the observer on the eastern side and towards us on the western side, consistent with a toroidal magnetic field in the disk and halo.
    \item[-] We detect a bubble-like structure extending about 3 kpc from the galactic plane to the southern halo in the total intensity map. This feature is also observed in the polarized intensity map and the H$\alpha$ image. The magnetic field orientations within the bubble-like structure are aligned with its direction of extent. A corresponding shell-like structure is observed in the north of the disk. This bipolar structure may be attributed to a galactic superwind driven by star formation feedback in the nuclear region.
\end{itemize}

Our work demonstrates that VLA S-band data are an effective probe to examine the physical conditions of radio halos in nearby galaxies. The combination of polarization observations and RM-synthesis enables mapping the 3D structures of galactic magnetic fields. Future works employing the combined S, C, and L-band data and radio observations at higher spatial resolution promise to impose tighter constraints on the magnetic field structure of the galactic halo and the associated CR propagation processes.

\section*{Acknowledgements}

We acknowledge research grants from the China Manned Space Project (the second-stage CSST science project: {\em Investigation of small-scale structures in galaxies and forecasting of observations}), the National Natural Science Foundation of China (No. 12273036), the Ministry of Science and Technology of China (National Key Program for Science and Technology Research and Development, No. 2023YFA1608100), and the support from Cyrus Chun Ying Tang Foundations.
J.T.L. acknowledges the financial support from the National Science Foundation of China (NSFC) through the grants 12273111, and also the science research grants from the China Manned Space Project. RJD and MS acknowledge support by Deutsche Forschungsgemeinschaft through SFB\,1491. Y.Y. acknowledges support from the National Natural Science Foundation of China through the grant 12203098. TW acknowledges financial support from the grant CEX2021-001131-S funded by MICIU/AEI/10.13039/501100011033, from the coordination of the participation in SKA-SPAIN, funded by the Ministry of Science, Innovation and Universities (MICIU).

\bibliography{citations}
\end{document}